\documentclass[showpacs,aps]{revtex4}
\usepackage{amssymb}
\usepackage{graphicx,verbatim}
\usepackage{epstopdf}
\usepackage{slashed}
\begin{document}

 \title{Resolving infrared singularities of QCD through the vertex paradigm}
\author{John M. Cornwall\footnote{Email:  Cornwall@physics.ucla.edu}}
\affiliation{Department of Physics and Astronomy, University of California, Los Angeles Ca 90095}

 \begin{abstract}
\pacs{11.15.Tk,11.15.Kc}
We furnish details and extensions for the vertex paradigm and several related ideas introduced in recent unpublished talks.  The vertex paradigm is a method for dealing non-perturbatively with the Schwinger-Dyson equations (SDE) of asymptotically-free (AF) gauge theories such as (for simplicity, matter-free) QCD, even in the face of necessary approximations.  It provides a useful truncation for the infinitely-many SDE of the gauge- and renormalization-group invariant Pinch Technique (or PT-RGI for short).  We implement the vertex paradigm by successive approximations at the one-dressed-loop level, postulating input tree-level gluon and ghost propagators as well as a 3-gluon vertex that are well-behaved in the infrared and also satisfy several crucial PT-RGI Ward identities that are QED-like and ghost-free.  Good IR behavior is assured by including a (non-running) gauge-invariant dynamical gluon and ghost mass as part of the input.  The non-trivial part of  the vertex paradigm is that, with our inputs, the one-loop output   vertex then satisfies the correct Ward identity from which we can construct the output gluon propagator, once we take proper account of Nambu-Goldstone-like massless scalars and related technical problems that arise whenever there is dynamical gluon mass generation.         The one-loop outputs show a number of desirable features: They are PT-RGI; free of any reference to a coupling (dimensional transmutation); satisfy the ghost-free Ward identities connecting them; give exactly the known one-loop UV behavior;  and are free of IR singularities.  We give a much simpler (because it is free of spin complications) illustration of the main principles of the vertex paradigm in a modified $\phi^3_6$ model that is AF.  Our successive-approximation scheme is not designed to estimate the gluon dynamical mass, but it shows that there is a lower limit to the mass below which the AF theory completely breaks down. 
     
\end{abstract}

\maketitle

\section{Introduction}

\subsection{General}

One of the major goals of analytic investigations of non-perturbative QCD is, or should be, to provide an approach for off-shell Green's functions and their SDE that:
\begin{enumerate}
\item  Is gauge-invariant
\item Satisfies the necessary Ward (or Slavnov-Taylor) identities
\item Is renormalization-group invariant (RGI)
\item Shows dimensional transmutation (their solutions do not depend on the coupling $g$)
\item Gives the exact and known UV behavior (asymptotic freedom, or AF)
\item Has no spurious  IR divergences coming from gauge artifacts \cite{abip,bib,hubsme}
\item Cures the IR singularities of perturbative QCD
\end{enumerate}

All of these criteria should be satisfied even given the unavoidable approximations needed in any approach to a $d=3,4$ AF gauge theory.  We could also add that the approach should be, in some context, simple enough to explain to graduate students who know the basics of quantum field theory.  

Criterion 1 is met by the Pinch Technique (PT), now more than thirty years old 
\cite{corn076,cornbinpap}.  So is criterion 2, in principle, but approximations could cause violations.  The point of the vertex paradigm is to find a non-perturbative approximation within the PT for vertices  that satisfies the Ward identities discussed in Sec.~\ref{wisec} and improves on the well-used gauge technique \cite{corn076,cornbinpap,corn090}.   A great number of papers have been written on the PT, some of which we will have occasion to cite later on.  

Although criteria 3, 4, and 5 are old and well-known, applications that respect them are more recent \cite{corn138,corn141,corn144}.  One may be surprised that a non-perturbative approach to an AF gauge theory fails to give the correct perturbative behavior in the UV, but in certain cases (including the original paper on the PT \cite{corn076}) it does fail.  The failure comes from particular truncations of the SDE, notably the gauge technique, which are more or less accurate in the IR but not in the UV.

Criterion 6 is a technical issue.  As pointed out in \cite{abip,bib,hubsme} in the Landau-gauge SDE the ghosts are massless, leading to spurious  IR divergences, but that does not occur in the PT, where ghosts have the same mass as gluons.

Criterion 7 has been the goal of many an analytic study.  The first idea \cite{corn076}, that AF forces the gluon to acquire a dynamical mass that is large enough to cure the IR singularities, is now widely accepted, in large part because lattice simulations in the Landau gauge (and occasionally in other gauges) show unequivocally that the
gluon propagator $D(p)$ is finite and non-zero at zero momentum \cite{alex1,alex2,bmmp,bouc1,bowm1,bouc2,silva1,silva2,bogo1,bogo3,stern,bogo2,born,
olivsilv,oliv,silvoliv,bowm2,ilge2,cucch1,cucchmend,cucch2,cucch3,bouc3,pawsp,zhang}.  These references investigate a variety of conditions, including different lattice sizes, finite temperature, Gribov copies, and the like. In all simulations the propagator is not diverging to infinity   near $p=0$, as in perturbation theory, nor does it approach zero or have a non-trivial power-law behavior there.  Although the Landau gauge is not the same as the PT, it is hard to escape the conclusion that the Landau-gauge propagator is showing some sort of dynamical gluon mass $m$, with $D^{-1}(p=0)\sim m^2$.  The Papavassiliou group has been engaged for years in approximately solving PT-inspired SDE whose solutions apply directly to Landau-gauge simulations, and even use Landau-gauge lattice results as input or constraints on their solutions \cite{agbinpap,papiban,binibanpap,aimp,agpap,pap,binpap08,binpap08-2,abp}; all these results show a dynamical gluon mass.  There are several other theoretical approaches to the Landau-gauge SDE {\cite{hubsme,sfk,pennwil,hubsme2,meyerswan}, all of which also find gluon mass generation.  

One obvious criterion is missing from our list:  The results of SDE studies should agree with ``experiment" (that is, lattice simulations).  But lattice simulations are not yet available for the PT, or equivalently the background-field method (BFM) Feynman gauge.

So far, no theoretical study or lattice simulation has been sufficiently powerful to determine the precise nature of this mass, which cannot be a simple pole mass, because gluons are held to other gluons by a breakable adjoint string.  This string   can and does break when the gluons are pulled sufficiently far apart, in exactly the same fashion as an unquenched quark-antiquark string does \cite{corn146}.  Just what the fact that $\widehat{\Delta}^{-1}(0)\neq 0$ means about the nature of the gluon mass is presently unclear.  Although it is not a conventional mass such as a free field may have,  we will model it as such.      String breaking is a non-perturbative effect that is not captured in any present-day approach to gluon mass generation, so such approaches are likely to generate a conventional mass for the gluon.  In this sense, the gluon mass has exactly the same foundations as the constituent quark mass does; in fact, both running masses behave like the VEV of a condensate, divided by $p^2$, at large momentum.  For gluons this is the usual $\langle G_{\mu\nu}^2\rangle $ condensate \cite{lavelle}, and that the running mass depends on it is interpreted \cite{corn076,cornbinpap} as supporting the reasoning that a condensate drives the dynamical mass, while a gluon with a dynamical mass leads to a condensate of quantum solitons \cite{cornbinpap} not found in the classical zero-mass theory.  Purists may argue that a mass that cannot be directly measured, as the proton mass can be, is no mass at all.  But there are other ways to get at the mass.  For example, in ordinary quantum mechanics it is certainly possible to extract $\omega$  from the energy levels of the potential energy $V=\omega^2x^2+\lambda x^4$ for the anharmonic oscillator, and there is a qualitative difference between the $\omega = 0$ case and the $\omega \neq 0$ case.  See \cite{corn146} for a discussion of gluonic string breaking and its relations to the dynamical gluon mass.

Many of these theoretical works truncate the SDE with something like the gauge technique \cite{corn076,corn090,pap}, in which the 3-gluon vertex is approximated by a function of the proper self-energy and, if necessary, terms involving ghosts.    But this is not necessarily an accurate approximation, and can explicitly fail to give the right UV behavior \cite{corn076,corn090}.  That is one reason why we study the vertex paradigm, in which the 3-gluon vertex is primary and the gluon propagator a secondary object, to be found (approximately) from the vertex.

\subsection{The vertex paradigm and our approximation to it}

The idea of the vertex paradigm is to start from a plausible input   3-gluon vertex and inverse propagator satisfying the criteria above.  The input 3-gluon vertex is free except for some terms, required by gauge invariance, that are proportional to $m^2$, the squared gluon mass, and the input propagator is a free massive propagator. Then we approximate the output vertex by inserting the inputs into the vertex PT-RGI SDE (in practice, at one-loop order).  Provided that  the necessarily approximate output vertex satisfies the QED-like Ward identity (see Sec.~\ref{wisec}) relating its divergence to the inverse propagator, we then get an output propagator.  As one might imagine, satisfying the Ward identity is non-trivial beyond perturbation theory, since some part of the input Green's functions must be non-perturbative in order to satisfy the last criterion:  No perturbative IR singularities coming from AF.  This scheme of successive approximations could be repeated, in principle, but it is extraordinarily complex beyond the simple one-loop ideas we use here.

We are not required to be completely systematic in any non-perturbative approach to QCD. The closest we can come seems to be to use the tree-level vertices and propagators from  adding to the usual NAGT action  a gauged non-linear sigma (GNLS) model action, multiplied by $m^2$, and using an appropriate gauge.  As is well-known, the GNLS action has local gauge invariance.  It is not obviously renormalizable, but that will not concern us at the present one-loop level of approximation.   In practice at the one-loop level, this amounts to using the usual tree-level 3- and 4-gluon vertices modified with $m^2$ corrections that come from the GNLS term, while the tree-level PT-RGI gluon propagator $\widehat{\Delta}$ and ghost propagator $\widehat{\Delta}_{gh}$ have the form
\begin{equation}
\label{treeprop}
\widehat{\Delta}_{\mu\nu}(p)=\frac{\delta_{\mu\nu}}{p^2+m^2}.\;\;\;\widehat{\Delta}_{gh}(p)=\frac{1}{p^2+m^2}.
\end{equation}
In the PT-RGI, the ghost mass is the same as the gluon mass.  In the Landau gauge the ghost is massless (more generally, the ghost mass depends on the gauge), which induces spurious IR singularities in the vertices and gluon propagator   that cannot occur in any gauge-invariant and physical quantity.  We discuss this in more detail in Sec.~\ref{vertpa} and Appendix \ref{gnlssec}.

Although the true dynamical gluon mass must run with momentum, it is critical to our approximation that the mass $m$ is constant, because then the Ward identity we need is satisfied at tree level for any {\em constant} mass.  This fact is essential in constructing an output 3-gluon vertex that also satisfies the same Ward identity.  We will argue in Appendix \ref{gnlssec} that there is a choice of gauge-fixing term yielding this simple massive propagator for an NAGT with a GNLS model mass term.  Of course, one can consider the case of a running mass, but the complexities defeat our purpose of as much simplicity as possible.

The fact that in our approximation the mass $m^2$ does not run   means that we cannot calculate it, even approximately, because certain integrals diverge logarithmically.  We can only remove the divergence by imposing a self-consistency condition on the results of the successive approximation at one loop, which is enough to show that not only must QCD generate a gluon mass dynamically, there is a lower limit $m_c$ to this mass, in terms of the QCD mass scale $\Lambda$, of $m_c\approx 0.66 \Lambda$.  In reality the mass must run or the SDE have no solutions.

Long ago, the PT one-loop 3-gluon vertex was given in perturbation theory \cite{corn099}, and shown to obey the QED-like Ward identity.  Sec.~\ref{vertpa} along with Appendices \ref{wipv} and   \ref{intpt} give arguments that the first---and biggest---step in doing this for the vertex paradigm is just to modify the results of \cite{corn099} by changing   {\em both gluon and ghost propagators  from massless to massive, as in the denominator of Eq.~(\ref{treeprop})}.    This gets us most of the way to satisfying the Ward identity, but by itself would be incomplete and inconsistent.  The steps needed to remedy these faults are outlined in Sec.~\ref{details} and discussed in more detail in  Sec.~\ref{vertpa} and Appendix \ref{intpt}.  All these steps generate output terms that have numerators linear in $m^2$.  Some terms  have massless longitudinally-coupled poles whose residue is $\sim m^2$; these are precisely the poles of the GNLS model.  They play the same role as  Nambu-Goldstone poles, although there is no symmetry ``breaking".  Even without symmetry breaking, such poles are required for any locally gauge-invariant mechanism for a dynamical mass for gauge bosons.

As one might expect, there are serious computational barriers for even the one-loop calculations in a gauge theory.  Therefore, we will give some details of a spinless Abelian model previously discussed briefly \cite{corn144} that is a special version of massless $\phi^3_6$, long known to be AF, in Sec.~\ref{phi36}.  In this model, all the criteria given above can be defined and are in fact fulfilled (except that there are no lattice simulations to compare with).  This   model has a certain {\em anschaulichkeit} (visualizability) because there are no spin complications.

\subsection{\label{details} Details and complications}

 In all orders of (massless) NAGT perturbation theory the   PT Green's functions are the same as those of the background-field method (BFM)  Feynman gauge. This is a distinguished gauge because there are no longitudinal momenta either in the tree-level propagator (which is the $m^2=0$ form of Eq.~(\ref{treeprop})) or in the tree-level vertex (which is $G^{0F}$ of Eq.~(\ref{fdecomp})). 
To oversimplify, the essence of the PT is to use these and other longitudinal momenta to modify ordinary Feynman graphs, but no such modification arises in the BFM Feynman gauge.     Beyond tree level the propagator does have longitudinal momenta in its numerator, and if mass is generated some of these terms necessarily have massless poles akin to those of Nambu-Goldstone fields.  Even at tree level the GNLS model leads to such terms, so the BFM Feynman gauge is not so simple.   In Appendix \ref{wipv} we note that the input propagator of Eq.~(\ref{treeprop})  comes from the 't Hooft-Fujikawa-Lee-Sanda \cite{thooft,fls} Feynman gauge with a special Higgs sector.      This gauge procedure was developed for NAGTs with Higgs fields and symmetry ``breaking", but the GNLS mass term is equivalent \cite{corn066} to $N$ Higgs-Kibble fields in the fundamental  representation of $SU(N)$) in the limit where the VEVs of the Higgs fields are frozen, and thee is no symmetry breaking.  Even though the Higgs fields are in the fundamental representation, there is a custodial symmetry that makes the massive gluons blind to the center symmetry of the gauge group.  In the FLS Feynman gauge,   tree-level vertices are those of the massless theory, plus an additional vertex that we discuss later.   

It is critical to note that in an FLS Feynman gauge the ghosts have the mass $m$ of the gluons, and in a more general FLS-$R_{\xi}$ gauge the ghost masses are gauge-dependent and unphysical.  Simply using this gauge for conventional Feynman graphs   does not realize the principles of the PT-RGI procedure, but when standard PT algorithms are used on the GNLS mass term with any gauge, the result is, as expected, PT-RGI and the ghosts do have the gluon mass.   
Generation of a dynamical gluon mass produces several complications:
\begin{enumerate}
\item  The massless poles mentioned above that appear in vertices and the propagator, and must satisfy the Ward identities by themselves.
\item Seagull graphs do not vanish as they do (with dimensional regularization) for massless gluons.  They are dealt with in  Sec.~\ref{masspart} by a regularization identity \cite{corn076} that does not use dimensional regularization.  This regularization has also been used often by Papavassiliou and collaborators.
\item Even though there are no explicit powers of $m$ in the non-pole Green's functions before momentum-space integration, such powers arise after this integration.  By themselves they can cause violations of Ward identities, all of which are cured with seagull terms as regulated in Sec.~\ref{masspart}.
\end{enumerate}

 We will see that using the one-dressed-loop PT-RGI equations, we get the exact UV behavior to one loop.  While this has not been investigated extensively for higher numbers of dressed loops in an NAGT, it is easy to work out in the tweaked $\phi^3_6$ model, where successive UV terms  of, say, the running charge that would be ordered by powers of $g^2$ in perturbation theory are ordered by the number of loops in the skeleton expansion.  Each skeleton graph has the UV behavior expected of an AF theory.

\section{\label{ptrgi}  The PT-RGI procedure}

  A renormalizable field theory (to be specific, in $d=4$) has a finite number of parameters (masses, couplings) and wave-function renormalization constants ($Z$ for short) that suffer infinite renormalization, so that these parameters are arbitrary, in the absence of further information outside the context of the given field theory. Because the $Z$s are infinite, the corresponding renormalized and finite Green's functions (proper self-energy, vertices) apparently necessarily depend on one or more renormalization schemes and renormalization masses $\mu$ that can be chosen almost arbitrarily.  The finite renormalized $S$-matrix is independent of such choices, but it has an undetermined finite parameter for every infinite renormalization constant.

  In QCD with massless quarks, there is only one mass parameter, the physical QCD scale $\Lambda$, and it does not appear in the classical action.   (Classical gluon and quark masses are forbidden by gauge and chiral symmetry respectively.   Any dynamically-generated masses are necessarily not only finite, but decrease at large momentum, typically like $1/p^2$ modulo logarithms.)   There is only one coupling constant $g$, and its arbitrariness is equivalent to the arbitrariness of $\Lambda$.  All dimensionless physical quantities, such as mass ratios, can be calculated unambiguously, including the running charge.  Yet the standard approach to renormalization in QCD is to introduce a renormalization mass $\mu$ and renormalization constant $Z_i$ for each irreducible Green's function needing renormalization (the 2-, 3-, and 4-point functions, including those with ghosts).   Each $Z_i$ is gauge-dependent, but certain products of the $Z_i$  are gauge-invariant.   

In the PT-RGI approach there is no need for this proliferation of renormalization constants.  In this section we review and give more detail for the PT-RGI procedure, in which for gluonic Green's functions we need only one $Z$ and no renormalization masses at all.  This circumstance leads directly to criteria 1-5 above.    It would be difficult  to  find equivalent results with conventional gauge-dependent Green's functions.

   Even in the usual PT (without the RGI feature), which yields gauge-invariant Green's functions, there is $\mu$-dependence.  We show that this dependence is removed by dividing the proper one-particle-irreducible  (1PI) PT functions by $g_0^2$, leaving Green's functions that are renormalization-group invariant (RGI), or PT-RGI Green's functions.  These are gauge-independent, process-independent, finite, and RGI.  All the Schwinger-Dyson equations (SDEs) for these Green's functions are manifestly independent of $\mu$ and of the coupling $g^2(\mu )$, as required by dimensional transmutation.  This $\mu$-independence is what would happen if an asymptotically-free NAGT were actually finite, as opposed to needing renormalization.  Dividing by $g_0^2$ is equivalent to finding the Green's functions of a quantum gauge potential that is the product of $g$ with the canonical gauge potential.

This holds for  Green's functions all of whose external legs are gluons,  but these gluonic Green's functions can have internal ghost or quark loops.  Ghosts cannot offer any obstacles, because one can always carry out the PT-RGI process in a ghost-free gauge; of course, the PT is independent of a gauge choice, and in such a gauge there are no unphysical Green's functions with external ghost legs to worry about.  As for quarks and ghosts, the $S$-matrix has no graph with an open quark or ghost line, and closed loops do not affect the PT-RGI property.  An RGI property holds for quark-gluon and ghost-gluon   closed loops.

There are two respects in which this development of PT-RGI Green's functions is important, beyond the usual virtues that PT Green's functions  are gauge- and process-independent.  The first is that, just as PT Green's functions can be made gauge- and process-independent even with various modes of approximation, PT-RGI Green's functions can be made $\mu$-independent even though these Green's functions are only approximately determined.  This is a step beyond construction of the $S$-matrix with conventional Feynman graphs, because when these are only approximate there is always some residual $\mu$-dependence.  The hope is that such dependence fades away as the approximation is improved (typically by going to more loops).     

The second respect is that if all off-shell Green's functions of QCD are finite and $\mu$-independent, with no arbitrary parameters at all, QCD begins to look like a finite field theory, even though it is really not finite.  Over the years since $S$-matrix theory was dominant in hadronic physics, there have been several clues that an NAGT such as QCD comes close to realizing the old $S$-matrix theory dreams of a finite theory \cite{corn141}.   

\subsection{Essence of the PT-RGI procedure}

In a general renormalizable field theory with cubic couplings, for any  three-particle vertex with bare coupling $g_0$ there are several renormalization constants:
\begin{equation}
\label{genvert}
g_0=\frac{Z_Vg_R}{(Z_1Z_2Z_3)^{1/2}},\;\;\Gamma_U=\frac{\Gamma_R}{Z_V}. 
\end{equation}
Symmetries of the theory may imply  relations between these $Z_i$, and in the PT construction 
 for the 2-, 3-, and 4-point proper gluonic PT-RGI Green's functions of a NAGT there is only one $Z$, for much the same reason that $Z_1=Z_2$ in QED:  There is a Ward identity constraint.   

Let us accept this for the moment, and write the renormalization relations between renormalized (R) and unrenormalized (U) PT Green's functions.  No matter what gauge is used, the standard PT propagator $d_{\mu\nu}$, three-gluon vertex $\Gamma_{\beta\alpha\rho}$, four-gluon vertex $\Gamma^{(4)}_{\beta\alpha\rho\mu}$, and bare coupling $g_0^2$ are renormalized by a single parameter $Z$, even though they may have internal quark or, in a more conventional gauge, ghost loops. 
\begin{equation}
\label{renormdef}
   d_U=Zd_R;\;\;\Gamma_U=\frac{\Gamma_R}{Z};\;\;\Gamma_U^{(4)}=\frac{\Gamma_R^{(4)}}{Z};\;\; g_0^2=\frac{g_R^2}{Z}.
\end{equation}
Here $U$ and $R$ stand for unrenormalized and renormalized, and
  we  always suppress the group indices. At this point we do not need space-time indices either.  The PT-RGI Green's functions are then defined as:
\begin{eqnarray}
\label{renormdef2} 
    \widehat{\Delta} =   g_0^2d_U = g_R^2d_R, \\ \nonumber
 G(p_i)  =  \frac{\Gamma_U(p_i)}{g_0^2}=\frac{\Gamma_R(p_i)}{g_R^2}\\ \nonumber
  G^{(4)} = \frac{\Gamma^{(4)}_U}{g_0^2} = \frac{\Gamma^{(4)}_R}{g_0^2}\\
\end{eqnarray}
We need not indicate whether $\widehat{\Delta}$, $G$, and $G^{(4)}$ are renormalized or not; they are all RGI, and from now on we drop the labels $U,R$.  

Note that these equations refer to Green's functions where all external legs are background, in the sense of the background field method \cite{cornbinpap}. We denote a background leg by B and a quantum leg by Q.  To construct all-B Green's functions we  need, for example, a BQQ three-gluon vertex.  An example is the gluon loop of Fig.~\ref{vgraph}, where the internal gluon lines are Q lines, and so the BBB vertex is composed of three BQQ vertices.

\subsubsection{A factorization of the PT-RGI propagator}

Eq.~(\ref{renormdef}) suggests \cite{corn138} the factorization
\begin{equation}
\label{hruneqn}
\widehat{\Delta}(p)=\bar{g}^2(p)H(p)
\end{equation}
where we interpret $\bar{g}$ as the running charge.  Although $\widehat{\Delta}$ is unique, its factorization is not.  We will {\em define}  $H(p)$ as a standard massive propagator with a running mass:
\begin{equation}
\label{hdefinition}
H(p)= \frac{1}{p^2+m^2(p)}
\end{equation}
which then defines the running charge through Eq.~(\ref{hruneqn}).  [In the present paper we use only a non-running mass.]
In the UV, where the mass can be dropped, this agrees with a standard definition of the running charge, and is useful in the IR.

All that is necessary for physics are the Green's functions, and not any particular factorization.  With the factorization that we use, we show \cite{corn144} in Sec.~\ref{running} how the running charge can be defined directly from the 3-gluon PT-RGI vertex.  (Ref.~\cite{bip} has related work.) 

There are two important points to be made about any such factorization:
\begin{itemize}
\item In perturbation theory, and we assume in full QCD, we expect that $\bar{g}^2(p)\sim 1/\ln p^2$ and $H(p)\sim 1/p^2$ at large momentum.  This means that the propagator 
$\widehat{\Delta} (p)$ decreases faster than $1/p^2$ and cannot satisfy the K\"allen-Lehmann representation with a positive  spectral function \cite{corn138}.
\item Because all physics resides in the propagator and not in its factorization, the widely-used running charge is not unique in the IR, because it depends on how $H(p)$ is defined.  This is apart from any UV non-uniqueness arising from renormalization scheme dependence.
\end{itemize}

These results on gluonic Green's functions still hold if closed chiral fermion and ghost loops are included.  Of course, including massive non-chiral fermions will require renormalization constants for the bare masses.

\subsubsection{\label{fermions}  Including ghosts and massless (chiral)  fermions}

We define, as with gluons, a factorization:
\begin{equation}
\label{fermfact}
S(p)=\frac{J(p)}{\tilde{Z}_2(p)},\;\;J(p)=\frac{1}{\mathrm{i}\slashed{p}+M(p)}
\end{equation}
Here $J(p)$ is taken to be RGI, and
\begin{equation}
\label{fermrenorm}
S_U(p)=Z_2S_R(p),\;\;\Gamma_{\mu U}=\frac{\Gamma_{\mu R}}{Z_1},\;\;\tilde{Z}_{2U}(p)
=\frac{\tilde{S}_{2R}(p)}{Z_2}.
\end{equation}
In the PT the Ward identity is QED-like:
\begin{equation}
\label{fermiwi}
\mathrm{i}q_{\alpha}\Gamma_{\alpha}(p,p+q,q)=S^{-1}(p+q)-S^{-1}(p)
\end{equation}
as in QED, so $Z_1=Z_2$.  [One quark momentum ($p$) is ingoing, the other is outgoing.]
This means that every quark loop coupled to gluons is RGI, because it has as many quark-gluon vertices as gluon propagators.  In particular, adding quark loops to the three-gluon vertex   does not change the RGI properties already discussed for that vertex.  The same argument holds for ghost loops.

The Ward identity also allows precisely the same results as for gluons:  In a quark loop the UV behavior of the vertex and the propagator cancel, so one can construct a useful quark-gluon vertex from a one-loop integral over free massive propagators, with both a CSB quark mass $M$ and a gluon mass $m$ to tame the IR behavior.

PT-RGI Green's functions satisfy ghost-free Ward identities that are the backbone of the vertex paradigm, and we take them up next.

\section{\label{wisec} The   Ward identities}

The 2-point Ward identity requires transversality of the gluon proper self-energy, and we
write, in an $R_{\xi}$ gauge,
\begin{equation}
\label{loopprop}  \widehat{\Delta}^{-1}_{\mu\nu}(k)=P_{\mu\nu}(k)\widehat{\Delta}^{-1}(k)+\frac{1}{\xi} k_{\mu}k_{\nu},\;\widehat{\Delta}^{-1}(k)=k^2+\widehat{\Pi}(k).
\end{equation}
where
\begin{equation}
\label{proj}
P_{\mu\nu}(k)=\delta_{\mu\nu}-\frac{k_{\mu}k_{\nu}}{k^2}
\end{equation}
is the transverse projector.
The scalar function $\widehat{\Delta}(k)$ is   gauge-invariant (independent of $\xi$), independent of the coupling $g$, and RGI.  

The all-order 3-gluon PT-RGI vertex Ward identity starts from the decomposition into a tree vertex and loop corrections:
\begin{equation}
\label{ptrgivert}
G_{\alpha\mu\nu}(q,k_1,k_2)=G^0_{\alpha\mu\nu}(q,k_1,k_2)+\widehat{\Lambda}_{\alpha\mu\nu}(q,k_1,k_2).
\end{equation} 
The loop corrections $\widehat{\Lambda}$ are unique, but the bare vertex comes in two forms.  The Ward identity for $\widehat{\Lambda}$ is:
\begin{equation}
\label{lambdawi}
q_{\alpha}\widehat{\Lambda}_{\alpha\mu\nu}(q,k_1,k_2)=P_{\mu\nu}(k_2)\widehat{\Pi}(k_2)
-P_{\mu\nu}(k_1)\widehat{\Pi}(k_1).
\end{equation}
If we choose the bare vertex as the standard NAGT vertex (divided by $g_0^2$), then
the Ward identity for the full vertex is:
\begin{equation}
\label{regwi}
q_{\alpha}G_{\alpha\mu\nu}(q,k_1,k_2)=P_{\mu\nu}(k_2)\widehat{\Delta}^{-1}(k_2)-P_{\mu\nu}(k_2)\widehat{\Delta}^{-1}(k_2).
\end{equation}

\subsection{Relation to background-field method vertices}

We will also use \cite{cornbinpap} the decomposition of the bare vertex $G^0$ into  parts, one of which is the BQQ vertex of the background-field method.  This decomposition singles out the background field momentum, which we call $q$.  It is simplest in the Feynman gauge ($\xi = 1$) that we will use throughout this paper, although it can be carried out for any gauge.  One part, called $G^P$,  contains all the pinches coming from longitudinal gluon momenta.  In Feynman gauge the remaining part is called $G^F$.    At tree level, with $G^0=G^{0F}+G^{0P}$:
\begin{eqnarray}
\label{fdecomp}
G^{0F}_{\alpha\mu\nu}(q,k_1,k_2) &  = & \frac{1}{g_0^2}[(k_1-k_2)_{\alpha}+2q_{\nu}\delta_{\alpha\mu}-2q_{\mu}\delta_{\alpha\nu}] \\ \nonumber
G^{0P}_{\alpha\mu\nu}(q,k_1,k_2) &  = & \frac{1}{g_0^2}[k_{2\nu}\delta_{\alpha\mu}-k_{1\mu}\delta_{\alpha\nu}]
\end{eqnarray}
We have:
\begin{equation}
\label{zerofwi}
q_{\alpha}G^{0F}_{\alpha\mu\nu}(q,k_1,k_2)=\frac{1}{g_0^2}(k_2^2-k_1^2)_{\mu\nu}
\end{equation}
which is the difference of inverse free propagators in the Feynman gauge, even the massive propagators of Eq.(\ref{treeprop}).  If we define
$G^F=G^{0F}+\widehat{\Lambda}$ the resulting Ward identity is:
\begin{equation}
\label{fullfwi}
q_{\alpha}G^F_{\alpha\mu\nu}(q,k_1,k_2)=  \widehat{\Delta}_{\mu\nu}^{-1}(k_2)-
\widehat{\Delta}_{\mu\nu}^{-1}(k_1)
\end{equation}
which follows from Eqs.~(\ref{loopprop}) with $\xi=1$ and (\ref{lambdawi}).  This is a QED-like Ward identity which will be critical in exploiting the vertex paradigm.

\subsection{\label{running} The running charge from the Ward identity for the 3-vertex}

For completeness we give here a result found in \cite{corn144}, for the
   pole-free part of the three-gluon vertex. Since the pole terms of the vertex and inverse propagator separately satisfy the Ward identity, we will drop them here.   As before, the running charge is a {\em defined} quantity, given for this paper by Eqs.~(\ref{hruneqn},\ref{hdefinition}).  

We assume
\begin{equation}
\label{zeromom}
\Delta_{\beta\gamma}^{-1}(q)  = \frac{(q^2+m^2)}{\bar{g}^2(q)}P_{\beta\gamma}(q)+ \dots
\end{equation}
which has a massless longitudinal pole with residue $\sim m^2$.  
Write the vertex as
\begin{equation}
\label{vertequ}
G_{\alpha\beta\gamma}(p,q,-p-q)=G^0_{\alpha\beta\gamma}G(p,q,-p-q)+\dots
\end{equation}
where
\begin{equation}
\label{bornvert}
G^0_{\alpha\beta\gamma}(p,q,-p-q)=(p-q)_{\gamma}\delta_{\alpha\beta}+(2q+p)_{\alpha}\delta_{\beta\gamma}
-(2p+q)_{\beta}\delta_{\alpha\gamma}.
\end{equation}
We are only interested in this kinematic structure, of the many different ones in the full vertex.  The omitted terms include longitudinal poles, so we can set $m^2=0$ in Eq.~(\ref{zeromom}).
Saving only the kinematical structure of Eq.~(\ref{bornvert}), the linear terms both on the left and right of the Ward identity have the kinematics
\begin{equation}
\label{linterms}
p_{\alpha}[2q_{\alpha}\delta_{\beta\gamma}-\delta_{\alpha\beta}q_{\gamma}-\delta_{\alpha\gamma}q_{\beta}].
\end{equation}
Equating coefficients at $p=0$ yields:
\begin{equation}
\label{lethm}
G(0,q,-q)=\bar{g}^{-2}(q).
\end{equation}

\section{Schematic SDE and dimensional transmutation}

In this section we review a less-common \cite{blee,corn112} form of SDE, in which the 3-vertex SDE has no bare term in the loop graphs and the vertex kernel is 2PI (two-particle irreducible) instead of 1PI; see Fig.~\ref{2pikern}. 
\begin{figure}
\begin{center}
\includegraphics[width=4in]{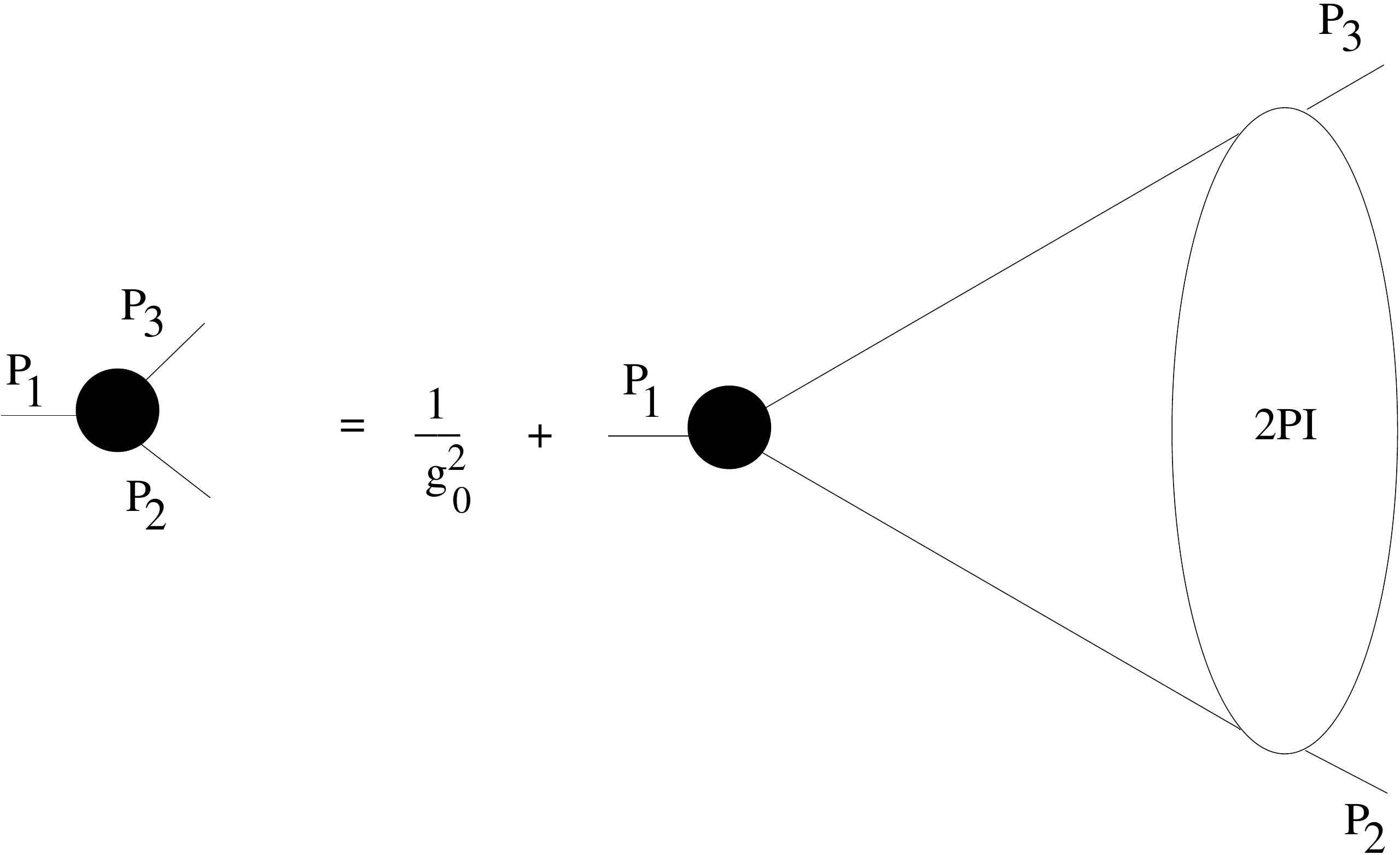}
\caption{\label{2pikern} A vertex graph with a 2PI kernel; all vertices and lines on the right-hand side are dressed.}
\end{center}
\end{figure}
\begin{figure}
\begin{center}
\includegraphics[width=5in]{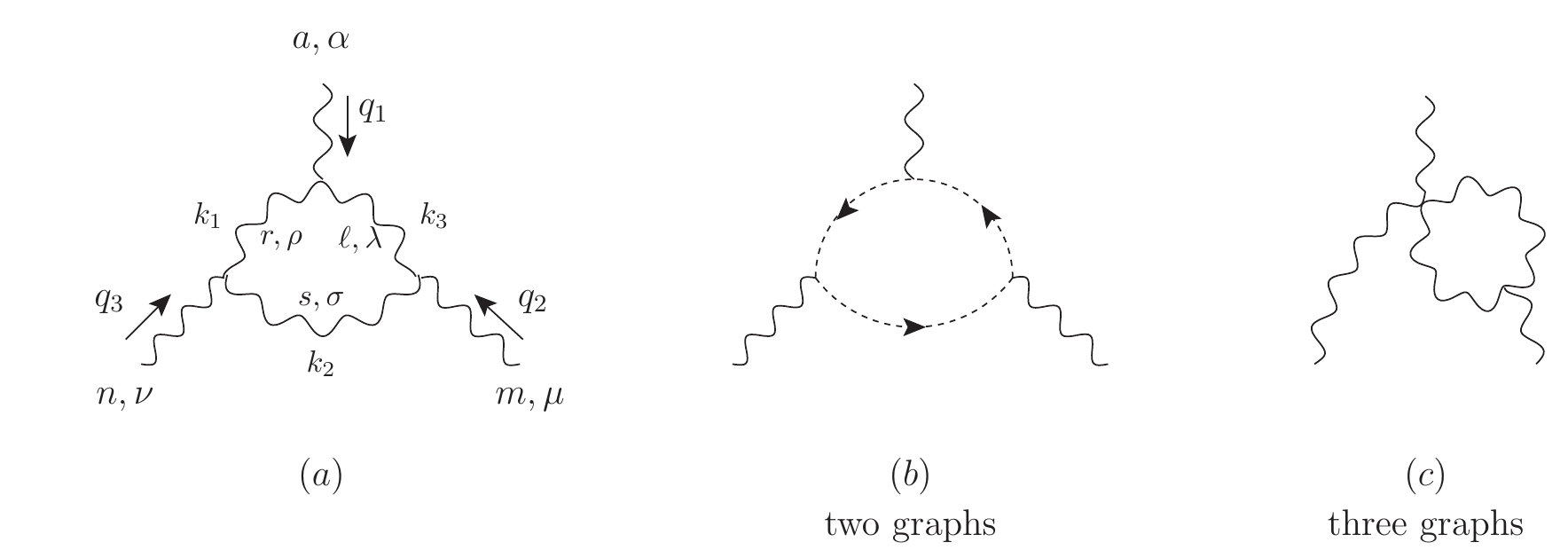}
\caption{\label{vgraph} The one-loop   graphs for     the non-pole part of the    three-gluon proper vertex.    [From \cite{cornbinpap}.]}
\end{center}
\end{figure}
One can interpret this equation as furnishing the bare vertex in terms of the dressed Green's functions, and eliminate the bare vertex wherever it appears in the usual SDE.
This has the possible disadvantage that the propagator SDE now has infinitely many terms, but since the ensemble of SDE must be truncated in any case, such a potential disadvantage is not a practical  hindrance.  The great advantage is that when such SDE are written for the PT-RGI Green's functions they are explicitly free of any reference to a coupling, which is replaced by reference to the physical QCD mass scale $\Lambda$.

We write the schematic (no group or space-time indices) 
SDE with just enough terms to understand what is going on.
For the one-gluon-loop of the propagator, and omitting numerical constants:
\begin{equation}
\label{propsde}
    \widehat{\Delta}^{-1}(p) =\frac{p^2}{g_0^2}+\int\!G^2\widehat{\Delta}^2+ \int\!G^{(4)}\widehat{\Delta} + \dots   
\end{equation}
Every vertex in the standard SDE has been divided by, and every propagator multiplied by  $g_0^2$, with the result that it disappears except in the Born term.  The omitted numerical constants are such that the final answer generates   $b$, the lowest-order coefficient in the beta-function. The (not displayed) ghost loop is RGI and independent of $g_0^2$ by itself, by the same argument that we give in Sec.~\ref{fermions}.  As far as RGI and dependence on $g_0^2$ goes, it is permissible to think of $G^{(4)}$ as equivalent to the product $G^2\widehat{\Delta}$, and a simple   argument shows that the one-loop propagator SDE depends only on the product $G\widehat{\Delta}$ (but an $N$-loop skeleton graph depends schematically on $(G\widehat{\Delta})^N\widehat{\Delta}^{N-1}$).
 
The one-loop BBB 3-vertex    comes from the graphs of Fig.~\ref{vgraph} (indices and constant factors suppressed).  
\begin{eqnarray}
\label{schverteq}
\Gamma_U & = & 1-g_0^2\int\!\mathrm{d}^4k\Gamma_U^3d_U^3+\dots \\ \nonumber
G & = & \frac{1}{g_0^2}-\int\!\mathrm{d}^4k\,G^3\Delta^3+\dots  
\end{eqnarray}
The second form (for $G$) of the equation needs no renormalization; it is actually finite.  This is because the divergence in $1/g_0^2$ (Eq.~(\ref{bareg}) below) cancels the divergence in the loop momentum integration---in fact, this cancellation can be used to find the form of the bare coupling as a function of a UV cutoff $\Lambda_{UV}$.
Of course, there are other graphs (see
Fig.~\ref{vgraph})   but it is enough for our understanding; the vertex will actually be based on the   evaluation \cite{corn099} of the perturbative PT vertex with a result equal to that of the BFM Feynman gauge.  These other graphs including 4-vertices, etc,  all lead to the property of independence of $\mu$ and $g^2(\mu )$.  

The difference between the bare and renormalized couplings, which at one-loop order are:
\begin{equation}
\label{bareg}
    \frac{1}{g_0^2}=b\ln (\frac{\Lambda_{UV}^2}{\Lambda^2})+\dots 
\end{equation} 
\begin{equation}
\label{renormg}
 \frac{1}{g^2(\mu )}=b\ln (\frac{\mu^2}{\Lambda^2})+\dots   
\end{equation}
is that the renormalized coupling depends on $\mu$ (and the physical scale $\Lambda$), while the bare coupling is independent of $\mu$.  So SDE such as Eq.~(\ref{schverteq}) are independent of $\mu$ also.

The Ward identity that follows from these SDE is:
\begin{equation}
\label{gluewi}   
p_{1\alpha}G_{\alpha\mu\nu}(p_1,p_2,p_3)=\Delta^{-1}(p_2)P_{\mu\nu}(p_2)-\Delta^{-1}(p_3)P_{\mu\nu}(p_3)  
\end{equation}
The Ward identity tells us that the product $\widehat{\Delta} G\sim 1/p^2$ at large $p$ and  the leading loop-order UV divergences are exactly those of the bare loop.    They will all cancel against the divergence(s) in $g_0^2$.
  Note that after cancellation this graph is the same as if the bare term were dropped and $\Lambda_{UV}$ replaced by the physical scale $\Lambda$.

We now have enough background to understand the whole PT-RGI procedure in the much simpler spinless model of the next section.  After that, we return to PT-RGI for QCD and the numerous complications there.

\section{\label{phi36} A tweaked $\phi^3_6$ model}

We begin \cite{corn141,corn144} with the much simpler SDE for the tweaked version of $\phi^3_6$, and then go on in Sec.~\ref{vertpa} to the $d=4$ gauge theory, which has many complexities coming from the gluon mass.  In what follows, think of the parameter $b$ as the one-loop coefficient of the gauge-theory beta function in $\beta = -bg^3+\dots$.  Because the NAGT has no bare mass, we omit such a term in the tweaked model that is supposed to mimic the NAGT.
\begin{figure}
\begin{center}
\includegraphics[width=4in]{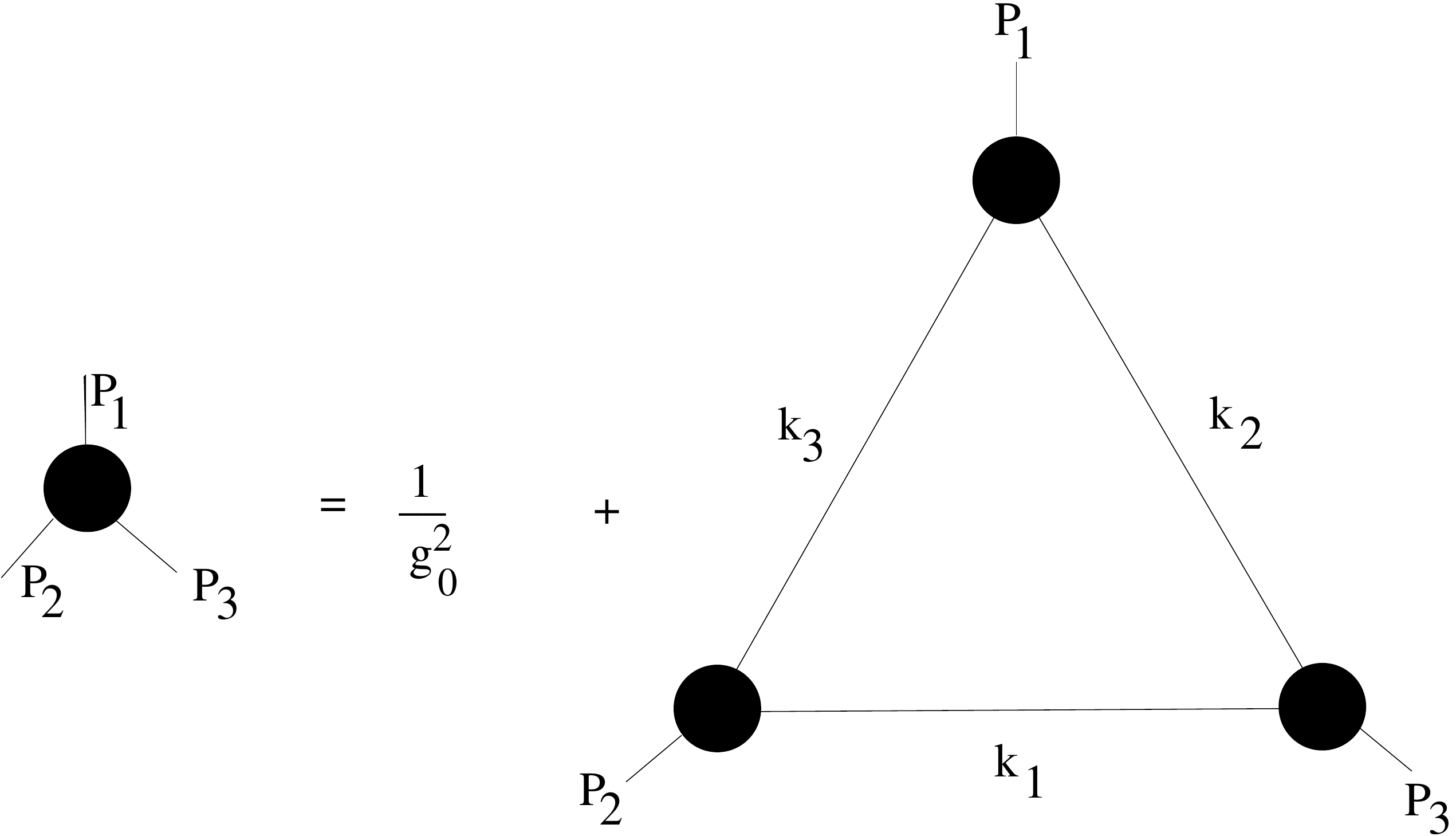}
\caption{\label{sde36} The one-loop skeleton graph for $\phi^3_6$ (from \cite{corn144}).  The $p_i$ are directed inward and the $k_i$ circulate clockwise. }
\end{center}
\end{figure}
The one-dressed-loop equations to solve for $\phi^3_6$ are (see Fig.~\ref{sde36}):
\begin{equation}
\label{phi361loop}
G(p_i)=\frac{1}{g_0^2}-\frac{2b}{\pi^3}\int\!\mathrm{d}^6k\,G(p_1,-k_2,k_3)G(p_2,-k_3,k_1)G(p_3,-k_1,k_2)
\prod_i\widehat{\Delta}(k_i) 
\end{equation}
There are no ghosts, and the Yang-Mills theory that our tweaked theory imitates has no expectation value $\langle A_{\mu} \rangle$, so we omit the tadpole graph.  The bare coupling is the same as given in Eq.~(\ref{bareg}), and as in the NAGT $\Lambda$ is the physical mass scale.

At the lowest order of the successive-approximation chain we use a simple propagator form with a constant mass $m$:
\begin{equation}
\label{simpleprop}
\widehat{\Delta} (p) = \frac{1}{p^2+m^2}.
\end{equation}
The starting $\phi^3$ vertex is just a constant, which we choose to be 1 so that, as required for an AF gauge theory, $G\widehat{\Delta} \rightarrow 1/p^2$ in the limit $p_i^2 \approx p^2 \gg m^2$.  

Now we add ``by hand" a Ward identity \cite{corn112,corn141}.  This allows us both to mimic the NAGT procedure and to avoid  the standard SDE for the propagator $\widehat{\Delta}$, which has a quadratic divergence not occurring in gauge theory. Consider two of the $\phi$ fields to be complex conjugates that carry an Abelian charge, and define \cite{corn112,corn141,corn144}    the propagator from the Ward identity for the (one-loop) SDE governing the  Abelian current.   The corresponding RGI-PT vertex $V_{\alpha}(p_i)$, with $p_1$ the momentum in the current-carrying vertex, obeys  a Ward identity analogous to Eq.~(\ref{regwi}):
\begin{equation}
\label{phi36wi}
p_{1\alpha}V_{\alpha}(p_1,p_2,p_3)=\widehat{\Delta}^{-1}(p_2)-\widehat{\Delta}^{-1}(p_3).
\end{equation}
By putting $p_2$ on shell, where the inverse propagator vanishes, we get $\widehat{\Delta}^{-1}(p_3)$.  Or, given the explicit form of the one-loop vertex, we can easily check that the Ward identity is satisfied, and after momentum-space integration the divergence reduces to a total derivative in Feynman parameters and is of the form (\ref{phi36wi}) for all momenta.  

At tree level the Ward identity is satisfied by the usual vertex
\begin{equation}
\label{abelvert}
V^0_{\alpha}(p_1,p_2,p_3)= (p_3-p_2)_{\alpha}.
\end{equation}
Beyond tree level, and again mimicking the NAGT case, we approximate the full vertex by multiplying $V^0$ by the same scalar vertex $G(p_i)$ occurring in the SDE for the scalar vertex, and omitting a term with kinematic coefficient $p_2+p_3$ that appears not to be important.    In the UV limit where all momenta scale with a momentum $p$ large compared to $m$ the Ward identity says that the product $G\widehat{\Delta}\approx 1/p^2$, clearly true at tree level.  The point here is that it is true even non-perturbatively, as we will now check using previous results \cite{corn141}.

 These results come from explicit integration of the one-loop scalar vertex:
\begin{equation}
\label{approxvert}
G(p_i)= \frac{1}{g_0^2}-b\int\![\mathrm{d}z]\ln [\frac{\Lambda_{UV}^2}{D+m^2}]
\end{equation} 
where 
\begin{eqnarray}
\label{zint}
\int\![\mathrm{d}z] & = & 2\int_0\!\mathrm{d}z_1\,\int_0\!\mathrm{d}z_2\,\int_0\!\mathrm{d}z_3
 \,\delta (1-\sum z_i),\\
D & = & p_1^2\,z_2z_3+p_2^2\,z_3z_1+p_3^2\,z_1z_2.
\end{eqnarray}
(The Feynman parameter $z_i$ goes with the line labeled $k_i$.)  
Using the form of $1/g_0^2$ in Eq.~(\ref{bareg}) we find a vertex that, in spite of being an approximation, is RGI and has no dependence on the coupling constant.  (One can also  scale out the coefficient $b$ of the one-loop equations, but we will not do that here.)
\begin{equation}
\label{approxvert2}
G(p_i)\approx b\int\![\mathrm{d}z]\ln [\frac{D+m^2}{\Lambda^2}].
\end{equation}
Similarly, the lowest-order form for the one-loop Abelian current vertex is
\begin{equation}
\label{photvert}
G_{\alpha}(p_i)=\frac{(p_2-p_3)_{\alpha}}{g_0^2}-3b\int\![\mathrm{d}z]\, \ln [\frac{\Lambda_{UV}^2}{D+m^2}]
 [p_2(1-2z_3)-p_3(1-2z_2)]_{\alpha}.
\end{equation}
in which the prefactor $3b$ is chosen to mimic gauge-theory results.  From the Ward identity of Eq.~(\ref{phi36wi}) we easily find the decomposition into two inverse propagators by observing that 
\begin{equation}
\label{qedward}
p_1\cdot [p_2(1-2z_3)-p_3(1-2z_2)]=[\frac{\partial}{\partial z_2}-\frac{\partial}{\partial z_3}]
[D+m^2].
\end{equation}
The Feynman-parameter integrals are trivial.  We do not add constants of integration that   would correspond to a bare mass, forbidden in the gauge theory we are mimicking, and find
\begin{equation}
\label{solvewi}
\widehat{\Delta}^{-1}(p)=6b\int_0^1\!\mathrm{d}z\,  [p^2z(1-z)+m^2]\ln [\frac{p^2z(1-z)+m^2}{e\Lambda^2}].
\end{equation}  
This (inverse) propagator has a standard kinetic term $\sim p^2$, a mass, and the usual momentum behavior for the $\phi^3_6$ one-loop propagator graph.  It differs from this graph only by specifying the kinetic and mass terms, amounting to specifying the arbitrary renormalization constants of the usual propagator graph.  But they are not arbitrary in a gauge theory,  so we use Eq.~(\ref{solvewi}) as it stands.

\subsection{Some new results in the tweaked model}

We ask several questions of the model propagator and vertex, designed to test whether the model looks much like what we expect from an AF gauge theory:
\begin{itemize}
\item While the correct leading UV behavior is guaranteed, how about the IR behavior?  In particular, to what extent does the inverse propagator of Eq.~(\ref{solvewi}) resemble the input $p^2+m^2$?
\item What does the running charge as defined earlier look like in the IR?
\item Does the one-loop propagator resemble the tree-level input propagator in the IR?
\item Do the  PT-RGI vertex and propagator from the tweaked model look like similar  approximations in the vertex paradigm of QCD?
\end{itemize}

One readily checks that one can always choose a value for $m/\Lambda$ such that $\Delta^{-1}(p^2)$  has a zero at $p^2=-m^2$; this value is $m/\Lambda = 1.802$.  This has the encouraging implication that $m/\Lambda$ has a reasonable value, and if for the sake of argument we put in the QCD-like value of $\Lambda \approx $0.3 GeV we get $m \approx $ 0.55 GeV, again a reasonable value that fits both lattice simulations and various theoretical estimates.  

Let us   expand $\widehat{\Delta}^{-1}(p^2)$ around $p^2=0$ from the integral in Eq.~(\ref{solvewi}).  The result is:
\begin{equation}
\label{numpropr}
\widehat{\Delta}^{-1} \approx b[1.070m^2+1.178p^2+\dots],
\end{equation}
not far from the input propagator (any overall factor such as $b$ is irrelevant).  Of course, there are deviations in the UV.

The next question is about the running charge.  Earlier we argued \cite{corn144}
that in an AF gauge theory one could define the inverse square of the running charge in terms of the coefficient of the Born term in the three-gluon vertex.  The analog in our model is:
\begin{equation}
\label{runchnum}
\bar{g}^{-2}(p)=G(0,p,-p)
\end{equation}
or at one loop, using Eq.~(\ref{approxvert2}):
\begin{equation}
\label{numval}
\bar{g}^{-2}(p)=b\int_0^1\!\mathrm{d}z\,2z\ln [\frac{p^2z(1-z)+m^2}{\Lambda^2}].
\end{equation}
In Fig.~\ref{runch} we plot $\alpha_s \equiv \bar{g}^2(p)/(4\pi )$ as a function of
$p^2/m^2$, using Eq.~(\ref{numval}) and the  value $b=11/(4\pi)$ for quarkless QCD.  In spite of the facts that our model is crude and bears no obvious resemblance to real QCD (aside from the value of $b$) the running charge is reasonably close to QCD expectations. 
\begin{figure}
\begin{center}
\includegraphics[width=4in]{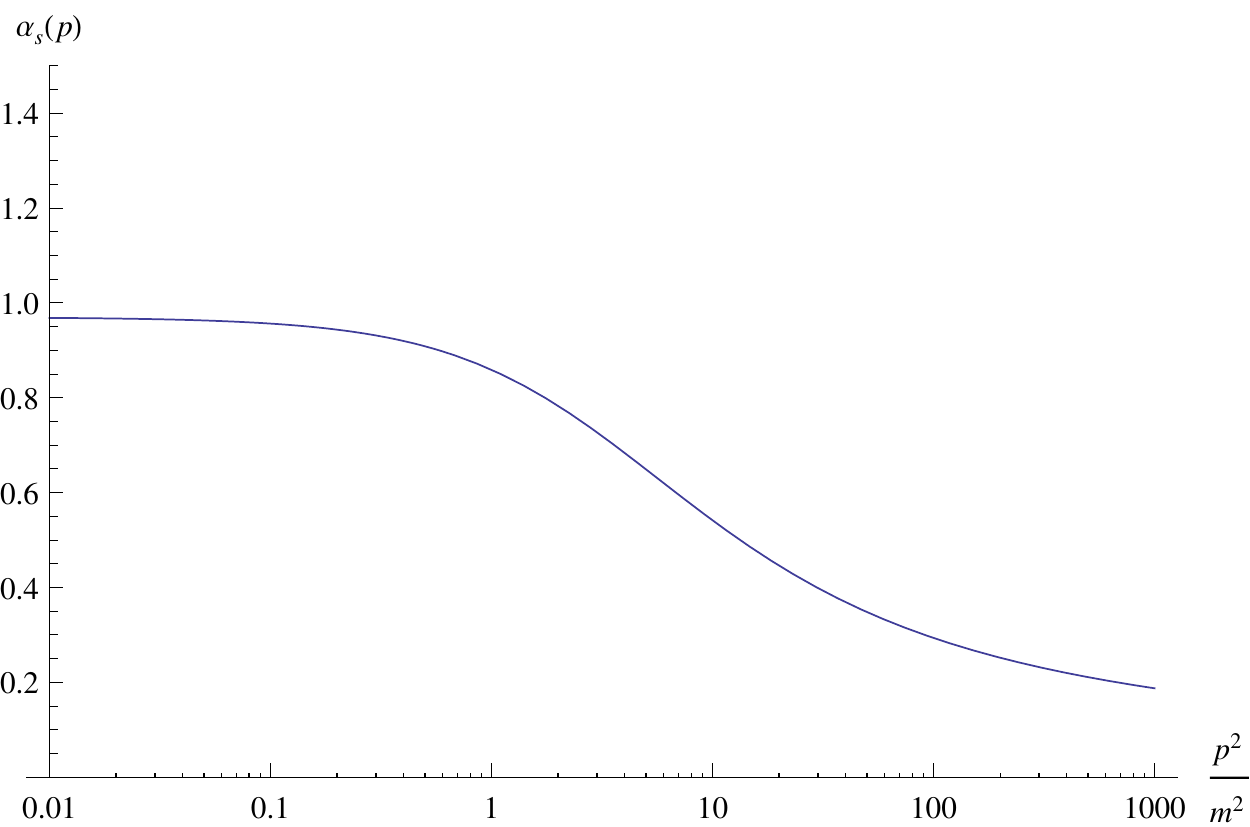}
\caption{\label{runch} A plot of $\alpha_s(p)$ as a function of $p^2/m^2$ with $m$ = 0.55 GeV.}
\end{center}
\end{figure}
In Fig.~\ref{deltaplot} we plot the dimensionless propagator $bm^2\Delta (p)$ using Eq.~(\ref{solvewi}), scaling out both the dimension and the factor $b$.
\begin{figure}
\begin{center}
\includegraphics[width=4in]{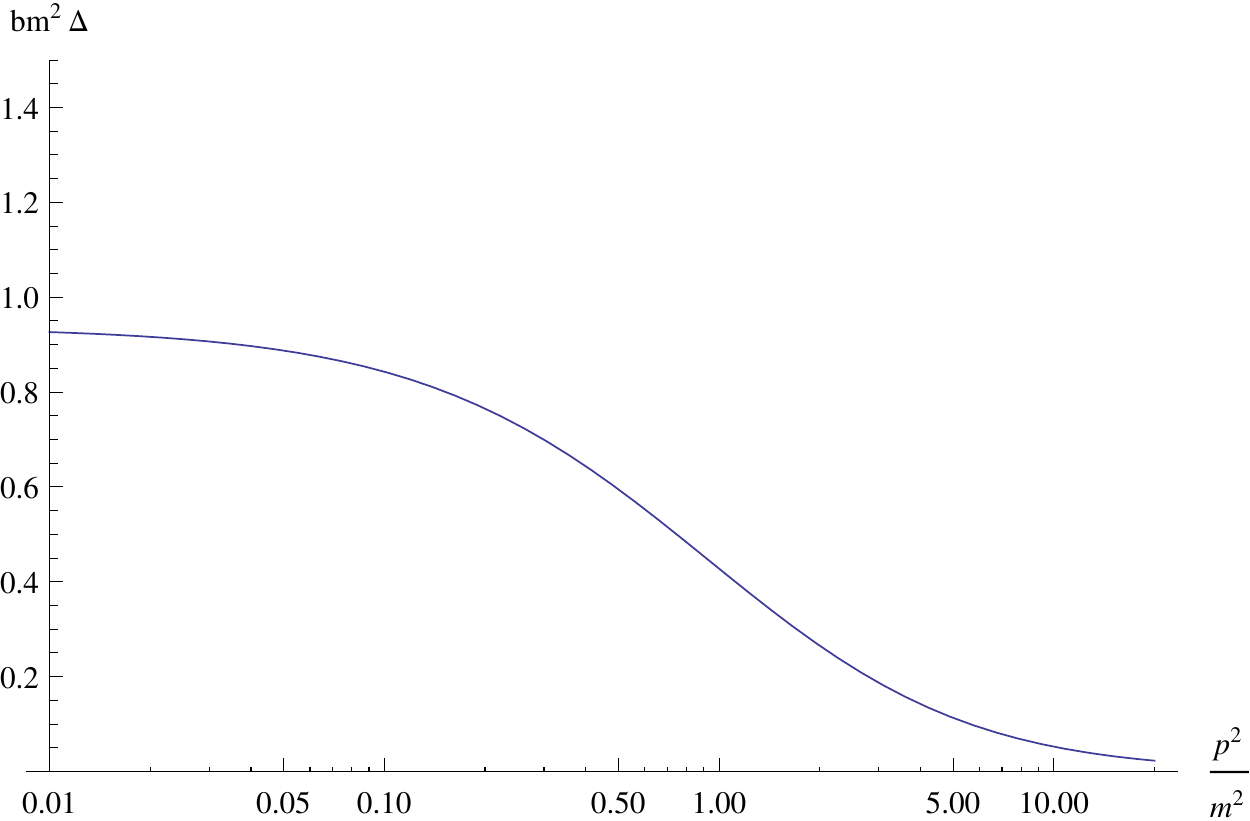}
\caption{\label{deltaplot} A plot of  $bm^2\Delta (p)$  as a function of $p^2/m^2$ with $m$ = 0.55 GeV.}
\end{center}
\end{figure}

\subsection{Higher orders}

  Figure \ref{2pi} shows the two-loop case.    
\begin{figure}
\begin{center}
\includegraphics[width=4in]{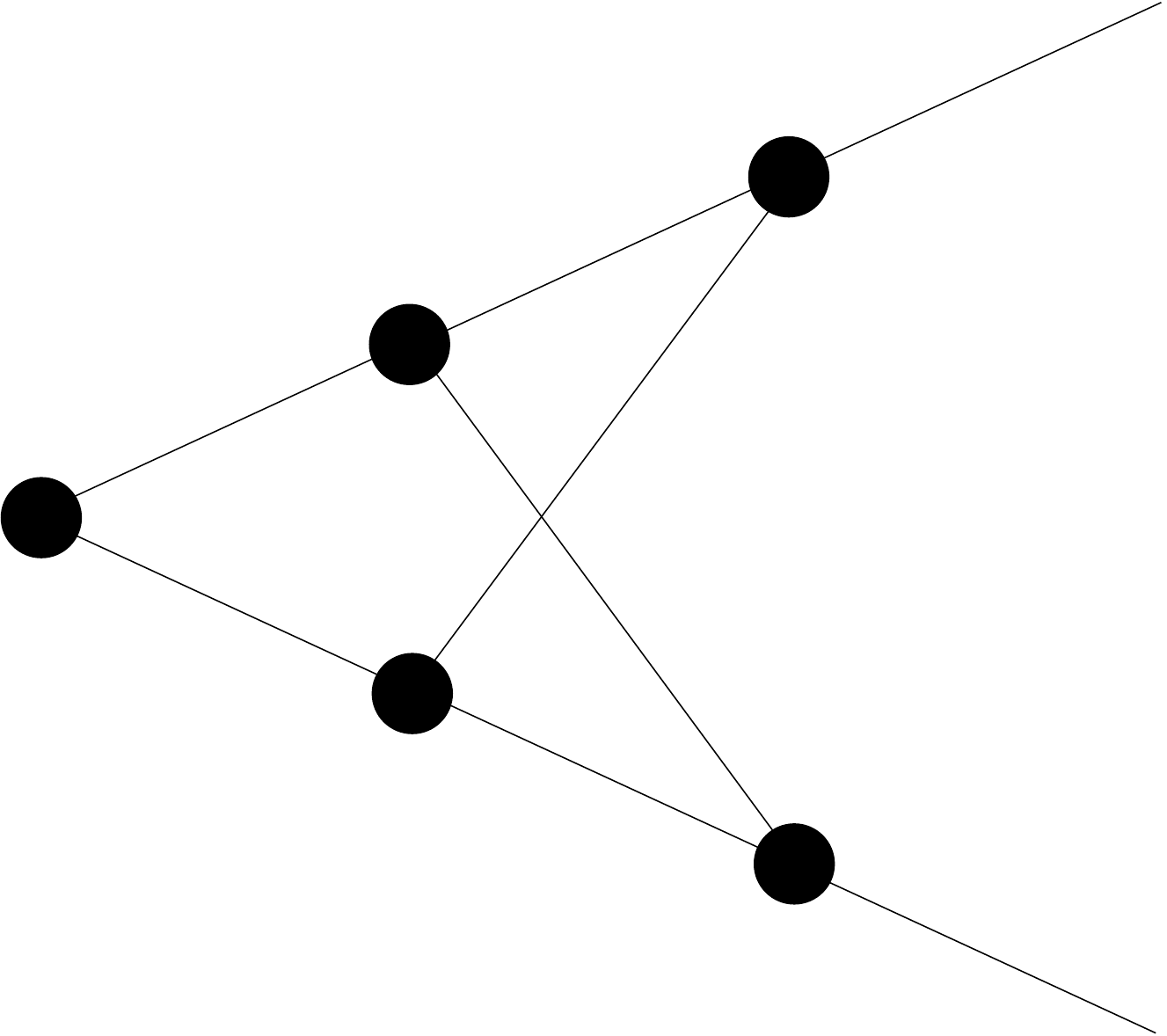}
\caption{\label{2pi} The two-loop 2PI skeleton graph for   the   three-gluon proper vertex, divided by $g^2$.     }
\end{center}
\end{figure}
Ref.~\cite{corn112} points out that every closed loop in any of  these graphs has at least four lines (the graphs are girth 4 in the graph-theoretic sense) and so have no Feynman parameter divergences.  The   momentum integrals have one overall logarithmic divergence (schematically, $\int\!\mathrm{d}k^2/k^2$).  Every loop added to a vertex graph adds three propagators, two vertices, and a power of $g_0^2$.  So at two loops the skeleton graph gives, schematically:
\begin{equation}
\label{highord}
\int\,\int\,g^2G^5\widehat{\Delta}^6
\end{equation}
which is an RGI combination; all higher order graphs are also RGI.  As \cite{corn112} shows, because of the extra propagators the UV behavior of the $N^{th}$ order graph comes from:
\begin{equation}
\label{highordint}
G_N\sim -\int^{\Lambda_{UV}^2}\!\mathrm{d}k^2\frac{1}{k^2 [\ln k^2]^{N-1}} 
\end{equation}
all the terms of which have the functional form expected from the RG for the inverse vertex.  The two-loop graphs gives $\sim \ln \ln p^2$.

The low-energy theorem of Appendix \ref{running} relates the vertex to $1/\bar{g}^2(p)$. One can show \cite{corn144} that to any finite order in $g$ only the first two terms of $1/\bar{g}^2(p)$ are growing at large momentum; all the rest are decreasing.  The first two terms are just those that 't Hooft points out \cite{thooft2} are RGI and universal.  
  With the correct UV behavior of the skeleton graphs and of the bare coupling,   UV logarithms cancel, effectively replacing $\Lambda_{UV}$ by $\Lambda$, the physical scale (NOT the renormalization scale).  All Green's functions are independent of $\mu$ and $g^2(\mu )$.

\section{\label{vertpa}  The vertex paradigm in QCD}

Now we turn to the really hard problem, the vertex paradigm in QCD.  
We want the simplest choice of input Green's functions satisfying the criteria of the Introduction.
From these we will calculate one-dressed-loop graphs for the three-gluon vertex,   as shown in Fig.~\ref{vgraph}.  The point here is that quite simple inputs can lead, even at only one dressed loop, to realistic output Green's functions satisfying all necessary WST identities.  The difficulties are in satisfying the basic Ward identity of Eq.~(\ref{regwi}); in dealing with the massless scalar excitations required for dynamical gluon mass generation; and in regularizing some seagull graphs that vanish at $m=0$.

 Here is a list of how we proceed; subsections and appendices take up the details.
\begin{enumerate}
\item To deal with IR singularities we change  the massless Feynman-gauge   tree propagator $ \delta_{\mu\nu}/p^2$ to $\widehat{\Delta}_{\mu\nu}(p)=\delta_{\mu\nu}/(p^2+m^2)$, where $m$ is a {\em constant} (non-running) mass.   We also take ghost lines to have exactly this mass.  There are gauges   where this can be enforced at tree level, and in any case the PT results have to be the same as in a ghost-free gauge.   But in the Landau gauge the ghosts stay massless even when the gluon gets mass.
\item In order to satisfy the critical Ward identity of Eq.~(\ref{regwi}), it is necessary that the mass $m$ is non-running.  With a constant mass we can satisfy various crucial Ward   identities without having to add mass terms in the numerators of graphs.  Therefore we require that all numerators of graphs are independent of $m$.   
\item Unfortunately, this makes it impossible to give a good estimate of $m$, because the SDE have a massive solution only if the mass vanishes at large momentum. See \cite{cornbinpap}. We are also unable to handle seagull graphs accurately, but we can add them by hand as necessary for satisfying Ward identities.  
\item   For the output vertex we calculate the usual one-loop BFM-Feynman gauge PT graphs \cite{corn099} for the three-vertex, but with massive propagators for gluons and ghosts, and exactly the same numerators as for the massless theory at one loop.  This output vertex, which we call $G^{(out)}$, can have no poles. 
\item To account for the massless poles that necessarily accompany mass generation, we add to $G^{(out)}$ the vertex $V^{gnls}$ of Eq.~(\ref{massv1}).  The Ward identity splits into two parts, with and without poles, and the WST for the pole vertex $V$ is precisely what is needed to satisfy the Ward identity (\ref{regwi}).  
\item Although the input vertices have no $m^2$ terms in the numerator, the output vertex does, even in the non-pole part.  These come from applying the PT to graphs such as in Fig.~\ref{sde} and from seagulls.  Those terms in the non-pole part that, after all integrations, are proportional to $m^2$ vanish, as they must by the WST identities, at zero momentum.
\end{enumerate}

\subsection{Satisfying the Ward identities}

Here we discuss the issues that distinguish the simple tweaked $\phi^3_6$ spinless model from real QCD.  Some of the complicated details we leave to Appendices.  The final result is a modest tweaking of the earlier calculation \cite{corn099} of the perturbative 3-gluon PT vertex at one-loop order, which  in fact yield the perturbative one-loop vertex in  the BFM-Feynman gauge \cite{cornbinpap}.The most important step to go from perturbation theory to the vertex paradigm is just to replace the usual gluon and ghost propagator denominators by $1/(p^2+m^2)$.  Note that  both the ghosts and the gluons have the same mass in the PT, while in, for example, the Landau gauge the ghosts stay massless even when the gluons are massive.  Fig.~\ref{vgraph} shows the one-loop graphs.

Of course, just putting massive propagators by hand is not a systematic approach to QCD.  But we are not required to be systematic, as long as we can satisfy the criteria listed in the Introduction.  As always in studying non-perturbative effects, the most important step is to start off with a decent approximation that can in principle be improved.  

The reason that putting in masses by hand is a good thing to do is that all the input vertices in Fig.~\ref{vgraph} individually satisfy ghost-free Ward identities that imply satisfaction of the overall Ward identity of Eq.~(\ref{regwi}) for the output vertex.     The 3-gluon vertices in this figure are all of $G^F$ type and obey the Ward identity in Eq.~(\ref{fullfwi}) as long as the gluon mass is non-running; the ghost vertices in the BFM Feynman gauge are convective and again satisfy a ghost-free Ward identity even with the gluon mass; and the 4-gluon tree-level vertex satisfies the Ward identity relating it to a sum of 3-gluon vertices independent of masses. 

In principle we could calculate explicitly the vertex just described, but it is a very lengthy result that is not yet done.  It is enough for us that this vertex, which we will call $G^{(out)}$, obeys a Ward identity from which we can write down the inverse propagator.

 However, if we were to stop here with just $G^{(out)}$ we would be wrong in an essential way.  The reason is that, by a Ward identity, the inverse propagator must have the form given in Eq.~(\ref{loopprop}), repeated here:
\begin{equation}
\label{loopprop2}  \widehat{\Delta}^{-1}_{\mu\nu}(k)=P_{\mu\nu}(k)\widehat{\Delta}^{-1}(k)+\frac{1}{\xi} k_{\mu}k_{\nu},\;\widehat{\Delta}^{-1}(k)=k^2+\widehat{\Pi}(k).
\end{equation}
If there is gluon mass generation then $\widehat{\Delta}^{-1} (k=0)\neq 0$ and there are inevitably massless Nambu-Goldstone-like poles in the inverse propagator.   And if there are such poles in the inverse propagator there must be corresponding poles in the PT-RGI vertex, by virtue of the Ward identity (\ref{regwi}).  But there is no possibility for a pole in the vertex part we constructed above, so something is missing.  We now turn to this problem, which actually was solved long ago \cite{corn090} in a different context.

\subsection{\label{vertpole}  Vertex parts with massless poles}

 First we discuss the constant-mass (or GNLS) case, and then the most general case.   The massless pole in the inverse propagator of Eq.~(\ref{invprop}) has  an $m^2$ residue and occurs in the right-hand side of the Ward identity (\ref{regwi}).  There must be terms in the left-hand side to balance,   and it does,   in a special vertex that we call $V$ that is to be added to the vertex part without poles.  From now on the terms $G,G^F$ will always refer to those pole-free parts, and the $V$ vertex is added to these. For the GNLS model we find \cite{corn090}:
\begin{equation}
\label{massv1}
V^{(gnls)}_{\alpha\beta\gamma}(p_1,p_2,p_3)=\frac{-m^2}{2}\{\frac{p_{1\alpha}p_{2\beta}}{p_1^2p_2^2}(p_1-p_2){\gamma}+ c.p.\}
\end{equation}
($c.p.$ means cyclic permutations) and it obeys the Ward identity
\begin{equation}
\label{v1wi}
p_{1\alpha}V^{(gnls)}_{\alpha\beta\gamma}(p_1,p_2,p_3)=\frac{m^2p_{2\beta}p_{2\gamma}}{p_2^2}
-\frac{m^2p_{3\beta}p_{2\gamma}}{p_3^2}
\end{equation}
which is $m^2$ times the difference of two longitudinal projectors.  

This vertex is a special case of one that holds for any (transverse) proper self-energy \cite{corn090}.  It is:
\begin{equation}
\label{corn090vert}
V_{\alpha\mu\nu}(p_1,p_2,p_3)= \frac{-p_{1\alpha}p_{2\mu}}{2p_1^2p_2^2}(p_1-p_2)_{\rho}
\Pi_{\rho\nu}(p_3)-\frac{p_{3\nu}}{p_3^2}[P_{\rho\alpha}(p_1)\Pi_{\rho\mu}(p_2)-
P_{\rho\mu}(p_2)\Pi_{\rho\alpha}(p_1)]+c.p.
\end{equation}
Add this to a pole-free vertex part, either of type $G_{\alpha\mu\nu}$ or $G^F_{\alpha\mu\nu}$ that obeys the pole-free part of the appropriate Ward identity, and the sum obeys the correct Ward identity with poles in the inverse propagator.  (If the scalar proper self-energy $\Pi(p^2)$ vanishes $\sim p^2$ at the origin, a lengthy calculation shows that the vertex $V$ has no poles.)  When one replaces the general proper self-energy by $P_{\rho\mu}(p\cdot )m^2$ in (\ref{corn090vert})  one gets the GNLS vertex of Eq.(\ref{massv1}).

Equations (\ref{massv1},\ref{v1wi}) say that the pole terms contribute additively both to the 3-vertex and to the inverse propagator, making it possible to write each as a sum of a term without poles and one with poles, such that the pole terms satisfy the Ward identity on their own.  So at this point the trial vertex is $G^{(out)}+V^{(gnls)}$.

But we are not yet finished.  There are two other ways that terms with $m^2$ in the numerator occur in the no-pole parts of the proper Green's functions.

\subsection{\label{masspart}  Induced-mass parts and seagulls}

Note that even though we did not put any powers of $m^2$ in graph numerators they appear nonetheless in the no-pole proper vertex and self-energy.  There are three sources of such terms:
\begin{enumerate}
\item Induced mass parts in the vertex and inverse propagator arising from using the so-called intrinsic PT \cite{corn099,cornbinpap}.  It is described in enough detail for our purposes in  Appendix \ref{intpt}.  In  the output propagator given in Eq.~(\ref{fullprop}), the induced mass term is the second term on the right-hand side of the first equation, after modification  with seagull terms as we describe momentarily.
\item Seagull graphs.  With our input propagator such graphs in the output should be of the form
\begin{equation}
\label{seagull1}
\delta_{\alpha\beta}\,\int\!\mathrm{d}^4k\,\frac{1}{k^2+m^2}.
\end{equation}
While this non-conserved integral vanishes at zero $m$, it does not for finite $m$.
The  seagull term can be rewritten using the identity \cite{corn076}:  
\begin{equation}
\label{intid}
\int \mathrm{d}^4k\,F(k^2)=-\int \mathrm{d}^4k\,[1+k^2\frac{\partial}{\partial k^2}]F(k^2)
\end{equation}
that removes power-law divergences and leaves convergent integrals unchanged.  This means that (for constant $m$) we can make the replacement
\begin{equation}
\label{massid}
\int\,\mathrm{d}^4k\,\frac{1}{k^2+m^2}\rightarrow -\int\,\mathrm{d}^4k\,\frac{m^2}{[k^2+m^2]^2}.
\end{equation}
It appears that this form of the seagull has a logarithmic divergence, but it is well-known that if a mass term is generated that has  no counterpart in the Lagrangian, this mass must run to zero at large momentum.  In fact, the mass runs as $1/k ^2$ (mod logarithms), and the seagull integral is convergent.
\item Terms $\sim m^2$ that appear after momentum integration in $G^{(out)}$.  Such terms in graphs with three propagators   are necessarily finite by power counting, but   two-line graphs with $m^2$ numerators in the vertex (Fig.~\ref{vgraph}(c)) can only give rise to seagull terms that we adjust by hand.  
\end{enumerate}

There are relations among these sources of $m^2$ numerators, as discussed in Sec.~\ref{outresult} below:
\begin{enumerate}
\item  All the terms in Eq.~(\ref{firstprop}) for the output inverse propagator, except the last, come from one-loop Feynman graphs where no poles can occur.  Moreover, the proper self-energy must be conserved.  But one easily checks that although   integrals such as $I_{\alpha\beta}(q)$, coming from these one-loop graphs, indeed has no poles it is not conserved.  The extra term necessary for conservation comes from seagull graphs, and  we will add it in by hand.  Adding it is equivalent to replacing $I_{\alpha\beta}(q)$ by 
$I_{\alpha\beta}(q)-I_{\alpha\beta}(0)$, or, for those terms in the proper self-energy with explicit factors of $m^2$, replacing $J(q)$ by $J(q)-J(0)$.  This has the further property that the momentum integral is convergent, as it must be (since there is no mass counterterm).
\item After adding seagulls as required, the only remaining logarithmic divergence is cancelled by the logarithmic divergence in $1/g_0^2$, just as in the tweaked $\phi^3_6$ model.  However, for the gauge theory there is no tweaking, because in the PT the one-loop beta-function coefficient $b=11N/(48\pi^2)$ is not an input, but an output of the calculation.  This means that the coefficient $b$ occurring in the one-loop bare charge of Eq.~(\ref{bareg}) is not to be thought of as an input either; in fact, the entire functional form of the bare charge is dictated by the PT-RGI SDE and is determined as the solution of these SDE progresses.
\end{enumerate}

\section{\label{outresult} Output results}

After carrying out much algebra that we do not give here, the procedures listed in Sec.~\ref{vertpa} finally yields the vertex-paradigm form of the one-dressed-loop PT-RGI inverse propagator:
\begin{equation}
\label{prop1}
\widehat{\Delta}^{-1}_{\alpha\beta}(q)=P_{\alpha\beta}(q)\frac{q^2}{g_0^2}+\widehat{\Pi}_{\alpha\beta}(q)+ g.f.t
\end{equation}
Here $g_0^2$ has the same form and interpretation as in  Eq.~(\ref{bareg}) and $\widehat{\Pi}_{\alpha\beta}$ is:
\begin{eqnarray}
\label{firstprop}
\widehat{\Pi}_{\alpha\beta}(q) & = & -\frac{N}{32\pi^4}\int \mathrm{d}^4k\,\frac{N_{\alpha\beta}}{(k^2+m^2)((k+q)^2+m^2)}\\ \nonumber
& - & \frac{2m^2N}{32\pi^4}P_{\alpha\beta}\int \mathrm{d}^4k\,\frac{1}{[k^2+m^2][(k+q)^2+m^2]}\\ \nonumber
 + \widehat{\Pi}^S_{\alpha\beta} +M^2P_{\alpha\beta}(q)
\end{eqnarray}
with
\begin{eqnarray}
\label{numerator}
N_{\alpha\beta} & = &\Gamma^F_{\alpha\mu\nu}\Gamma^F_{\beta\mu\nu}-2(2k+q)_{\alpha}(2k+q)_{\beta}\\ \nonumber & = & 8q^2P_{\alpha\beta}(q) + 2(2k+q)_{\alpha}(2k+q)_{\beta},
\end{eqnarray}
and $\widehat{\Pi}^S_{\alpha\beta}$ is a momentum-independent seagull that subtracts off $\widehat{\Pi}_{\alpha\beta}(0)$.  The $M^2$ term comes from the $V$-vertex.

Do  the index algebra to reduce   the second momentum integral in Eq.~(\ref{firstprop}) to scalar form:
\begin{eqnarray}
\label{integral2}
I_{\alpha\beta}\equiv \frac{1}{\pi^2}\int\!\mathrm{d}^4k\,\frac{(2k+p)_{\mu}(2k+p)_{\nu}}{(k^2+m^2)((k+p)^2+m^2)} & = &  ~\\ \nonumber 
 -\frac{1}{3}P_{\alpha\beta}(q)q^2J(q;\Lambda_{UV})-\frac{4}{3}\delta_{\alpha\beta}m^2J(q;\Lambda_{UV})-\frac{2}{3}m^2\delta_{\alpha\beta}J(0;\Lambda_{UV})
\end{eqnarray}
with
\begin{equation}
\label{jdef}
J(q;\Lambda_{UV})=\frac{1}{\pi^2}\int\!\mathrm{d}^4k\,\frac{1}{[k^2+m^2][(k+q)^2+m^2]}.
\end{equation}
It is understood that the $k$ integral is cut off at $\Lambda_{UV}$:
\begin{equation}
\label{juv}
J(q;\Lambda_{UV}) = -\int_0^1\!\mathrm{d}x\,\{1+\ln [\frac{m^2+x(1-x)q^2}{\Lambda_{UV}^2}]\}.
\end{equation}

Observe that even though before the momentum integration in Eq.~(\ref{integral2}) there is no $m^2$ in the numerator, after this integration such a term appears.  This term is logarithmically divergent, and could give rise to a pole in the one-loop integral for the self-energy, but this is impossible because $I_{\alpha\beta}$ comes from the Ward identity for  a vertex without such poles.  Furthermore, the integral $I_{\alpha\beta}$ is not conserved, as it should be.  

It is easy to see that addition of a seagull term that replaces, in this term, $J(q;\Lambda_{UV})$ by 
$J(q;\Lambda_{UV})-J(q=0;\Lambda_{UV})$ simultaneously solves all these problems:  The logarithmic divergence disappears, $I_{\alpha\beta}$ is conserved, and there is no pole in this term.
With the choices for seagulls given above, the PT-RGI proper self-energy is conserved, and therefore proportional to $P_{\alpha\beta}(q)$.    After the momentum integral the result for the scalar part of the inverse PT-RGI propagator, using the vertex paradigm, is: 
\begin{eqnarray}
\label{fullprop}
 \widehat{\Delta}(q)^{-1}  = -b[q^2J(q;\Lambda) - \frac{m^2}{11}[J(q;m)- J(0;m)]+ M^2 =\\
 b\int_0^1\,\mathrm{d}x\,\{q^2[1+\ln [\frac{m^2+q^2x(1-x)}{\Lambda^2}] ]+ 
\frac{m^2}{11}\ln [1+\frac{q^2 x(1-x)}{m^2}]\}+M^2
\end{eqnarray}
where $J(q;\rho)$ for any mass $\rho$ comes from substituting $\rho$ for $\Lambda_{UV}$ in Eq.~(\ref{juv}).  There are no divergences in this expression, and no pole in the $m^2/11$ term, because the coefficient vanishes at $q=0$.  

It is important to note that, unlike in the tweaked $\phi^3_6$ model, nowhere have we input the value of the beta-function coefficient $b=11N/(48\pi^2)$. This critical gauge-invariant parameter emerges from the calculation.   Our final result is virtually identical (with a redefinition of the seagull) to a previous result found using a different truncation, but the same input propagator \cite{corn138}.

It only remains to determine $M^2$.  Impose the condition that $\widehat{\Delta}^{-1}(q^2=-m^2)$ vanish, so that $m^2$ is the ``physical" mass of the gluon.  This leads to:
\begin{equation}
\label{meq}
M^2=bm^2\{\ln [\frac{em^2}{\Lambda^2}]-\frac{10}{11}(\frac{\pi}{\sqrt{3}}-2)+1\}\approx bm^2\ln [\frac{2.3m^2}{\Lambda^2}].
\end{equation}

\subsection{The running charge}

The running charge $\bar{g}^2(q)$ is one of the most often-invoked concepts in QCD. But {\em in principle} it is actually redundant; every pure-glue QCD process can be described in terms of the PT-RGI propagator and vertices without every mentioning the running charge or any other kind of coupling.  Nevertheless in practice it is a useful concept, and in Sec.~\ref{ptrgi} we suggested the product decomposition of Eq.~(\ref{hruneqn}):  $\widehat{\Delta}(q)=\bar{g}^2(q)H(q)$, with $H$ given by the free massive propagator of Eq.~(\ref{hdefinition}).   This has the trivial non-uniqueness of multiplying $\bar{g}^2$ by any constant and dividing $H$ by the same constant; we resolve this at large momentum by choosing the perturbative form  $H\rightarrow 1/q^2$.   Then from the  large-$q$ form of $\widehat{\Delta}$ in Eq.~(\ref{fullprop}):  
\begin{equation}
\label{propuv}
\widehat{\Delta}(q)\rightarrow \frac{1}{bq^2\ln [\frac{q^2}{\Lambda^2}]}=\frac{\bar{g}^2(q)}{q^2}
\end{equation}
the remaining factor is indeed the leading term of  the usual running charge.  

Just as in Refs.~\cite{corn076,corn138}, the spectral form of the propagator (\ref{propuv}) in Minkowski space has a  non-positive spectral function, as all lattice simulations show (see \cite{corn144} for a review and references).  This is because $\widehat{\Delta}(q)$ decreases faster than $1/q^2$ at large momentum.

Our suggested form for $H$ in the IR is the massive input propagator $H=1/(q^2+m^2)$, although we certainly cannot prove that this is uniquely determined by any known QCD physics. It is often used in phenomenology at very low momentum, such as near-forward diffractive processes. A different   $H$ gives a different form for $\bar{g}^2$ in the IR, but this is a matter of choice as long as the choices made do not change $\widehat{\Delta}$ itself.  From Eq.(\ref{fullprop}) we find:
\begin{equation}
\label{irprop}
\widehat{\Delta}(q=0)\approx\frac{1}{bm^2\ln [\frac{2.3m^2}{\Lambda^2}]}.
\end{equation}
Then our choice of $H$ gives:
\begin{equation}
\label{runchir}
\bar{g}^2(0)=\frac{1}{b\ln [\frac{2.3m^2}{\Lambda^2}]}.
\end{equation}
If $m=2\Lambda$, as various estimates suggest, then   $\alpha_s(0)\approx 0.51$ for $SU(3)$ with no quarks, and perhaps 0.6 for three quark families.  This is in good agreement with an analysis \cite{courtoy,courtoy2} of experimental data on the Bjorken sum rule, which yields $0.42\leq \alpha_s(0)\leq 0.58$, with a central value of 0.5.
A simple interpolation formula, useful for {\em spacelike} (Euclidean) momentum, between the UV and the IR is:
\begin{equation}
\label{interpol}
\bar{g}^2(q)\approx \frac{1}{b\ln [\frac{q^2+2.3m^2}{\Lambda^2}]}.
\end{equation}
This form, with 2.3 replaced by 4, was suggested in the first paper \cite{corn076} on the PT.  Observe that for $m < 0.659\Lambda$ the propagator has a tachyonic pole and is unphysical; this means that dynamical mass generation is necessary to make an AF gauge theory a sensible field theory.

Fig.~\ref{dplot} shows the Euclidean propagator, times $bm^2$, for the value
$m/\Lambda =2$.  Except for an overall normalization factor this propagator is remarkably similar to that of the tweaked $\phi^3_6$ model, shown in Fig.~\ref{deltaplot}.  There is a sense in which all AF pure-glue theories look alike.
\begin{figure}
\begin{center}
\includegraphics[width=4in]{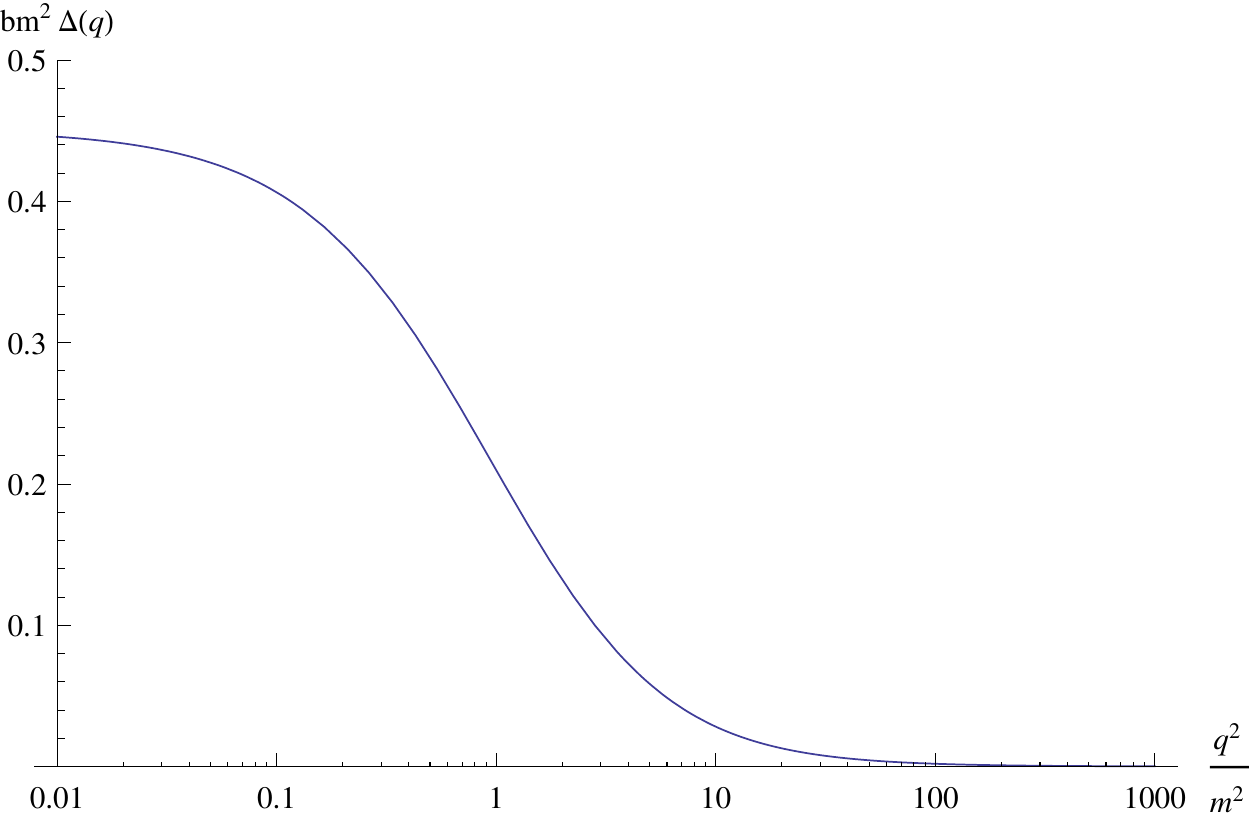}
\caption{\label{dplot} A plot of the PT-RGI propagator times $bm^2$, for $m/\Lambda = 2$.}
\end{center}
\end{figure}

Fig.~\ref{yvert} shows the pole-free 3-vertex form factor $G(p,-p,0)$ for $SU(3)$, which according to \cite{corn144} and Sec.~\ref{running} should be the same as $1/\bar{g}^2(p)$.   
\begin{figure}
\begin{center}
\includegraphics[width=4in]{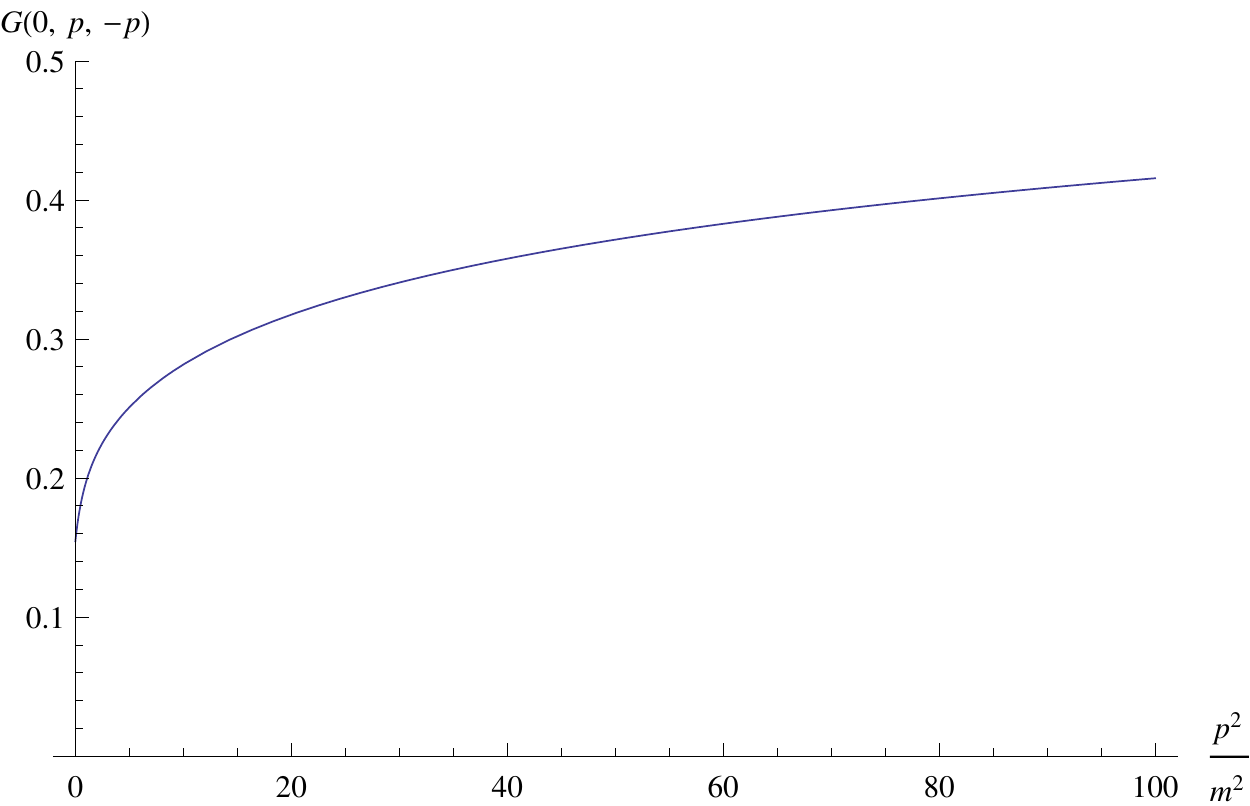}
\caption{\label{yvert} A plot of the PT-RGI 3-vertex form factor $G(p,-p,0)\equiv 1/\bar{g}^2(p)$, for $SU(3)$.}
\end{center}
\end{figure} 
(For other $N$ with gauge group $SU(N)$, multiply $G(p,-p,0)$ by $N/3$.)
As it happens, a closely-related quantity has been studied on the lattice \cite{cmm}:  A certain projection of the Landau-gauge 3-vertex.  This projection removes all longitudinal momenta and therefore removes the entire pole vertex.  In the Landau gauge the ghosts have zero mass, so that the ghost loop gives a logarithmically-divergent contribution, as in perturbation theory.  The lattice data of \cite{cmm} are not sufficiently accurate in the IR to resolve this divergence, and in the UV the Landau-gauge 3-vertex will not rise exactly as $\ln p^2$, as does the PT-RGI form factor, but as a different power.  For related work, see \cite{bip}.

\section{Final remarks}

We have given a simple implementation of a new method, called the vertex paradigm, for dealing with non-perturbative problems of off-shell gluon Green's functions in QCD, based on a first approximation in which the gluon has a non-running mass $m$.  The procedure is at every step tied to the PT-RGI process, guaranteeing gauge invariance, renormalization group invariance, dimensional transmutation, as well as that all Green's functions are independent of the S-matrix process in which they occur.  The paradigm constructs from this input, and tree-level input vertices, a 3-gluon output proper vertex   that satisfies a number of desirable criteria listed in the Introduction.  One of these criteria is a QED-like (ghost-free) Ward identity that yields an estimate of the gluon proper self-energy from the output vertex.

Although the current implementation is too simple to estimate $m$, other more elaborate procedures suggest that $m\approx 2\Lambda$, where $\Lambda$ is the physical QCD mass scale.  We estimate the corresponding zero-momentum coupling $\alpha_s(0)\approx 0.5$ (for no matter fields).  We are  able to estimate a minimum mass $m_c\approx 0.6 \Lambda$, and for $m\leq m_c$ there is a tachyonic pole and QCD breaks down.  

There is a sharp contrast between the PT-RGI methods used here and other methods, such as calculating \cite{hubsme} or simulating \cite{cmm} off-shell Green's functions in Landau gauge.  For one thing, in Landau gauge the ghost mass is zero even when the gluon has mass, but in our PT-RGI techniques both ghost and gluon must have the same mass. A zero ghost mass   in Landau gauge leads to unpleasant IR singularities.  For another thing, when one does not pay attention to the requirements of dimensional transmutation (that trades the coupling for a mass scale) one may inappropriately conclude that the running coupling of QCD is, or can be assumed to be, a process-dependent quantity.  See, for example, \cite{dbck} and references therein, which analyze the same data on the Bjorken sum rule that were used in \cite{courtoy}, but adopt the process-dependent normalization $\alpha_s(0)=\pi$. In perturbation theory (large momentum) the running coupling surely should be the same for all processes, and it makes sense to define it, as we do, to be the same for all processes no matter what the momentum is.

The main question for future applications of the vertex paradigm is accounting for the running of the dynamical gluon mass.  It might be possible to combine the methods of Papavassiliou {\em et al.} \cite{agbinpap,papiban,binibanpap} with the present work.  These authors construct an SDE approach that provides an equation for the running mass $m(p^2)$, the kernel of which depends on (among other things) gluonic vertices that are approximated with the gauge technique.  It should be possible instead to use a one-loop approximation along the lines of the vertex paradigm in this equation, as the first step in a successive-approximation chain.  This is far from an easy step, and I have nothing to report on it.

\newpage 
\appendix

\section{\label{gnlssec} The gauged non-linear sigma model}

While it was simple enough in $\phi^3_6$ to use a tree-level propagator with a mass, it is not so straightforward  in an NAGT.  One way that preserves local gauge invariance at the tree level (and for our application at the one-loop level, at least) is to add a GNLS term to the usual NAGT action.  This yields the total action
\begin{equation}
\label{gnls}
S=\int \mathrm{d}^4x\,\frac{1}{g^2}\mathrm{Tr}[\frac{1}{4}G^2_{\mu\nu}
+\frac{m^2}{2}(U^{-1}\mathcal{D}_{\mu}U)^2]
\end{equation}
where $U$ is, for $SU(N)$, an $N$-dimensional unitary matrix and $\mathcal{D}_{\mu}$ is the covariant derivative.  It turns out that the equations of motion for $U$ are simply the (identically vanishing) covariant derivative of the NAGT equations of motion.  These can be solved in perturbation theory, writing $U=\exp \omega$ and expanding in a series.
The lowest-order solution is
\begin{equation}
\label{lowgnls}
\omega = \frac{1}{\Box}\partial \cdot A+ \dots
\end{equation}
At higher orders one finds  more massless pole terms, including a new part to the 3-point vertex (and also the 4-point tree vertex that we will not need here) that has longitudinally-coupled massless poles.  For the 3-point vertex these are just the PT poles in Sec.~(\ref{vertpole}) above.

A gauge-fixing term needs to be added in order to define the propagator.  For example, we could use a standard $R_{\xi}$ term:
\begin{equation}
\label{sgf}
S\rightarrow S+S_{gf};\;\ S_{gf}=\int \mathrm{d}^4x\,\frac{1}{2\xi g^2}\mathrm{Tr}(\partial \cdot A)^2
\end{equation}
but these leads to complicated algebra involving cancellations between massless ghosts and the massless poles of Eq.~(\ref{lowgnls}) (see Eq.~(\ref{bareprop}) below).

\subsection{\label{wipv}  The Ward identity and Green's functions with Nambu-Goldstone-like poles}

Consider now the propagator decomposition of Eq.~(\ref{loopprop}) and the Ward identity of Eq.~(\ref{regwi}).  In an NAGT with no gluon mass generation the scalar proper self-energy $\widehat{\Pi}(p)$ has a factor $p^2$ at any order, and so there are no longitudinal pole singularities in $\widehat{\Delta}^{-1}$.  But if there is mass generation this is signaled by a finite value for $\widehat{\Pi}(p=0)$ and there is a pole term. In the case of the GNLS mass term in the Feynman gauge, the tree-level inverse propagator is:
\begin{equation}
\label{invprop}
\widehat{\Delta}^{-1}_{\alpha\beta}(p)=P_{\alpha\beta}(p)(p^2+m^2)+p_{\alpha}p_{\beta} 
= \delta_{\alpha\beta}(p^2+m^2)-m^2\frac{p_{\alpha}p_{\beta}}{p^2}
\end{equation}
corresponding to the propagator
\begin{equation}
\label{bareprop}
     \widehat{\Delta}_{\alpha\beta}(p)=\delta_{\alpha\beta} \frac{1}{p^2+m^2} +\frac{p_{\alpha}p_{\beta}}{p^2} [\frac{1}{p^2}-\frac{1}{p^2+m^2}].
\end{equation}
The massless pole in the inverse propagator is  essentially a Nambu-Goldstone pole, although there is no symmetry ``breaking" and no explicit Higgs fields.  But however gluon mass is generated in an NAGT, if local gauge invariance is preserved there must be such poles. In the PT such massless poles are important in cancelling massless ghosts in PT Green's functions, but just like Nambu-Goldstone poles they never appear in the S-matrix.

The form of the term in square brackets of Eq.~(\ref{bareprop}) as a difference of a massless and a massive term suggests a cancellation mechanism in which the first term is cancelled by the usual massless ghost, and then the massless ghost is reinstated with the same mass as the gluon.  Then we can interpret the result as producing a massive tree-level propagator in the usual Feynman gauge:
\begin{equation}
\label{massprop}  \widehat{\Delta}^{-1}_{\mu\nu}(p)= \delta_{\mu\nu}(p^2+m^2).
\end{equation}
accompanied by a ghost field of the same mass.
As in the massless case not having longitudinal terms in the propagator saves a lot of work in the PT.

Beyond tree level,
because the proper self-energy $\widehat{\Pi}_{\mu\nu}$ is conserved we need a gauge-fixing term such that this equation becomes
\begin{equation}
\label{rightprop}  \widehat{\Delta}^{-1}_{\mu\nu}(p)=P_{\mu\nu}(p)(p^2+\widehat{\Pi}(p^2))+g.f.t.
\end{equation}
Here $g.f.t.$ is the gauge-fixing term, proportional to $p_{\mu}p_{\nu}$, that is generated by loops. In our successive-approximation scheme the self-energy is just $m^2$, and one easily sees that    the simple form of Eq.~(\ref{massprop})  corresponds to
\begin{equation}
\label{gftform} \widehat{\Delta}^{-1}_{\mu\nu}(p)=P_{\mu\nu}(p)(p^2+m^2)+p_{\mu}p_{\nu}(1+\frac{m^2}{p^2}). 
\end{equation}

\subsection{The Fujikawa-Lee-Sanda gauge}

It may not be immediately obvious, but
this gauge-fixing term is what arises when the 't Hooft and Fujikawa-Lee-Sanda (FLS) principles of gauge fixing of a gauge theory with Higgs fields \cite{thooft,fls} are applied to a gauge theory with a GNLS term.  This works because (as noted in Sec.~\ref{details}) the GNLS term is equivalent \cite{corn066} to a Higgs-Kibble symmetry-breaking term of a special type, consisting of (for $SU(N)$) $N$ scalars $\psi_i$ in the fundamental representation, along with a custodial (global) symmetry that corresponds to gluons' transforming in the trivial representation of the center symmetry of the gauge group.  The $i^{th}$ component of the scalar $\psi_i$ is given a frozen VEV of $m$, and the matrix $U$ of the GNLS is given by $m^2U=\sum_i\psi_i\bar{\psi}_i$, followed by an $SU(N)$ rotation on the scalars.   The non-local term $m^2/p^2$ comes from replacing the Higgs-Kibble fields by their lowest-order expression in terms of the gauge potential, given in Eq.~(\ref{lowgnls}).  Presumably using the BFM for the FLS in Feynman gauge is equivalent to the PT, although this remains to be proved.

The point of the FLS gauge is to cancel the mixing term between the gradient of the Higgs-Kibble field and the gauge potential, with the result that, in the FLS Feynman gauge that we are using, the ghost has the mass of the gluon.  One can also see this directly from the GNLS model in the usual Feynman gauge, where the inverse propagator is given in Eq.~(\ref{invprop}).  The clue to this is the cancellation arising from the longitudinal terms in the propagator, which, when combined with all the other longitudinal (pinch) terms, does two things.  One is to replace the massless ghost by a   ghost with the gluon mass, and the other is to change the ghost-gluon vertices to the convective ones of the BFM Feynman gauge.

\section{\label{intpt} General rules for intrinsic pinching in FLS gauge}

The intrinsic pinch \cite{corn099,cornbinpap} does not need embedding of a Green's function in an S-matrix element,   but shortcuts the original PT algorithm by recognizing that the PT amounts to dropping inverse propagators referring to background (B) lines from the conventional Feynman Green's function carrying such lines.

   First we discuss the one-dressed-loop 3-vertex graph of Fig.~\ref{sde}, and then give a little more detail for the proper self-energy.  This discussion is not intended to be complete, but only to give some idea of the needed details.  To avoid a proliferation of factors of $1/g^2$ we discuss only the PT procedure; extension to PT-RGI is trivial.

\subsection{The intrinsic pinch and the 3-gluon vertex}

To set the vertex kinematics see Fig.~\ref{sde}.  We begin with input dressed vertices that have no massless poles, in which case no such poles can emerge from the integrations that yield the output vertex.  At the lowest level of successive approximation these ``dressed" vertices are simply tree-level vertices.  Pole terms are added as described in Sec.~\ref{vertpole}.
\begin{figure}
\begin{center}
\includegraphics[width=4in]{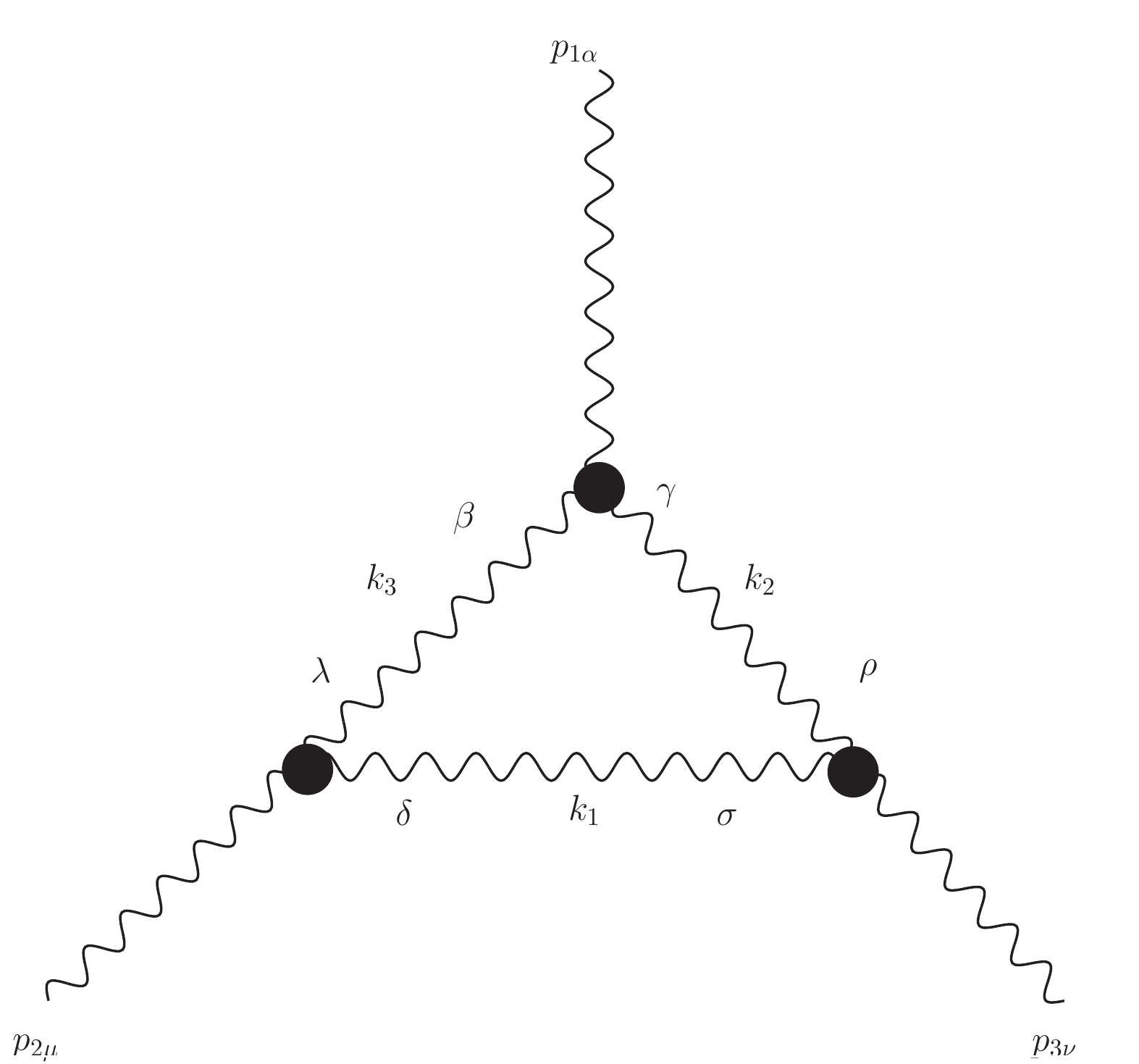}
\caption{\label{sde} The one-loop skeleton graph with three propagators for   the   three-gluon proper vertex.     All $p_i$ go into vertices and the $k_i$ go clockwise, so $p_1=k_2-k_3,p_2=k_3-k_1,p_3=k_1-k_2$.}
\end{center}
\end{figure}
For the propagator lines use Eq.~(\ref{treeprop}) as well as ghost propagators with the same mass; all vertices are the usual tree-level ones.
The first rule is to expand any product of conventional tree-level vertices $\Gamma$ in powers of $\Gamma^P$.  The way to do this is to write a product of $\Gamma^F$ in terms of $\Gamma -\Gamma^P$:    $\Gamma^F\Gamma^F\Gamma^F=
(\Gamma-\Gamma^P)(\Gamma-\Gamma^P)(\Gamma-\Gamma^P)$, and then express the original product of $\Gamma$ in the graph from this equation.
Pinches leading to inverse propagators come from Ward identities; these can be external (B line) pinches or internal pinches (Q lines).  

Provided that $m^2$ is constant in momentum space, and the same for all gauge bosons, then before pinching:
\begin{enumerate}
\item Because the vertices are the usual ones, all numerators are exactly the same as in massless Feynman gauge.   
\item This implies that the external-line Ward identity is the same as for the massless 
BFM Feynman gauge:
\begin{equation}
\label{usuwi}
p_{1\alpha}\Gamma^F_{\alpha\mu\nu}(p_1,p_2,p_3)=\delta_{\mu\nu}[(p_3^2+m^2)-(p_2^2+m^2)]=
\widehat{\Delta}^{-1}_{\mu\nu}(p_3)-\widehat{\Delta}^{-1}_{\mu\nu}(p_2)
\end{equation}
with the $p_{\mu}p_{\nu}$ terms missing for $\Gamma^F$.  The $m^2$ terms cancel.
This is the same, in terms of $\widehat{\Delta}$, whether or not this propagator has a   mass term.
\item The products  $\Gamma\Gamma\Gamma^P$ with one pinch factor yield the generic form 
\begin{equation}
\label{onegammap}
    -P_{\mu\nu}(p_i^2)p_i^2+P_{\mu\nu}(k_j)k_j^2
\end{equation} 
with one background propagator (momentum $p_i$)   and one quantum propagator.
The instructions of the intrinsic pinch are to drop  the background propagator;  any terms of the form $p_{i\mu}p_{i\nu}$ can always be dropped for background lines.  We   get, after dropping the B propagator, 
\begin{equation}
\label{postpinch}
   \delta_{\mu\nu}(k_j^2+m^2)-k_{j\mu}k_{j\nu} =\widehat{\Delta}^{-1}_{\mu\nu}(k_j)-k_{j\mu}k_{j\nu}
\end{equation}   
---expressed in terms of propagators, the same as in the massless case.  
 
\item The product  of two or more $\Gamma^P$s induces $m^2$ terms in numerators after dropping B inverse propagators. 
For example:
\begin{equation}
\label{pinchmass}
k_{1\sigma}\Gamma_{\sigma\mu\beta}(k_1,p_2,-k_3)=(k_3^2+m^2)\delta_{\mu\beta}-k_{3\mu}k_{3\nu}-
\textrm{pinch term}
\end{equation}
where the pinch term has a term $m^2p_{2\mu}p_{2\nu}$.  This is the same structure as the massive case, with a term generating a   vertex contribution with two gluon lines plus a term that adds to the ghost numerator. 
 
\item Products such as $\Gamma^P\Gamma^P$ and $\Gamma\Gamma^P\Gamma^P$ have terms $\pm k_i^2\delta_{\mu\nu}$ to which we must add and subtract $m^2\delta_{\mu\nu}$; this also generates a numerator $\sim m^2$ as above:
\begin{equation}
\label{pinchmass2}
-\Gamma^P_{\beta\alpha\rho}\Gamma^P_{\rho\nu\sigma}=-k_2^2\delta_{\alpha\beta}\delta_{\nu\sigma}+\dots
\end{equation}
which generates an induced term $m^2\delta_{\alpha\beta}$ when we complete the $k_2^2$ propagator.
\end{enumerate}

So we just use the   massless PT vertex, as given in Eq.~(3.18) of Ref.~\cite{corn099}, but changing both ghost and gluon propagator denominators to the massive form.  To this we add the pole vertices,    the induced $m^2$ term, and seagulls chosen to enforce transversality where required.  There is another source of terms with $m^2$ in the numerator beside the induced mass terms discussed above; other terms arise from doing momentum-space integrations and from seagulls.  See Sec.~\ref{masspart}.

The fact that the Ward identities with mass are so close in structure to the massless ones is what allows the 3-vertex as outlined above to satisfy its Ward identity, given in Eq.~(\ref{lambdawi}), and therefore yield the (approximate) propagator $\widehat{\Delta}$.
Some added work is needed to account for the induced mass terms, seagulls, and pole vertices, but we will not go into that here.  It is important to note that the induced mass terms of the can lead to the appearance of poles in vertices such as described here that are not allowed to have poles; one avoids such spurious poles by subtracting seagulls whose value is chosen just for that purpose.  

A summary of steps taken in practice, to implement the above discussion:
\begin{enumerate}
\item  Take the expression for the perturbative PT 3-gluon vertex, as given in Eqs.~(2.24,3.18) of Ref.~\cite{corn099}, and add $m^2$ to all massless propagator denominators.
This guarantees the Ward identity of Eq.(\ref{lambdawi}), because the massless Ward identity for $\Gamma^{0F}$ in Eq.~(\ref{zerofwi}) is also satisfied with constant masses.
Moreover, in the vertex of \cite{corn099}, as in the BFM method, the ghost vertices are modified to be convective, and their Ward identity is  the same as for $\Gamma^{0F}$ (with vector indices removed)---satisfied with or without constant mass terms.
\item Add a pole vertex (Sec.~\ref{vertpole}). The Ward identity (\ref{lambdawi}) is still satisfied.
\item Find the induced-mass terms arising from the intrinsic PT (Sec.~\ref{masspart}).
\item Find other terms linear in $m^2$ from momentum-space integrals and seagulls.  
\end{enumerate}

There is no point in writing the very lengthy form of the 3-gluon vertex explicitly.  Instead, we give only the result for the inverse PT-RGI propagator that follows from the vertex paradigm.  It is actually essentially identical with an inverse propagator  based on the gauge technique \cite{corn138}; this happens because both methods use the simple gluon propagator  of Eq.~(\ref{treeprop}).

\subsection{The intrinsic PT for the propagator with mass}

We use the kinematics of Fig.~\ref{propgraph} plus the corresponding ghost loop; seagulls are added ``by hand" as needed for Ward identities.
\begin{figure}
\begin{center}
\includegraphics[width=4in]{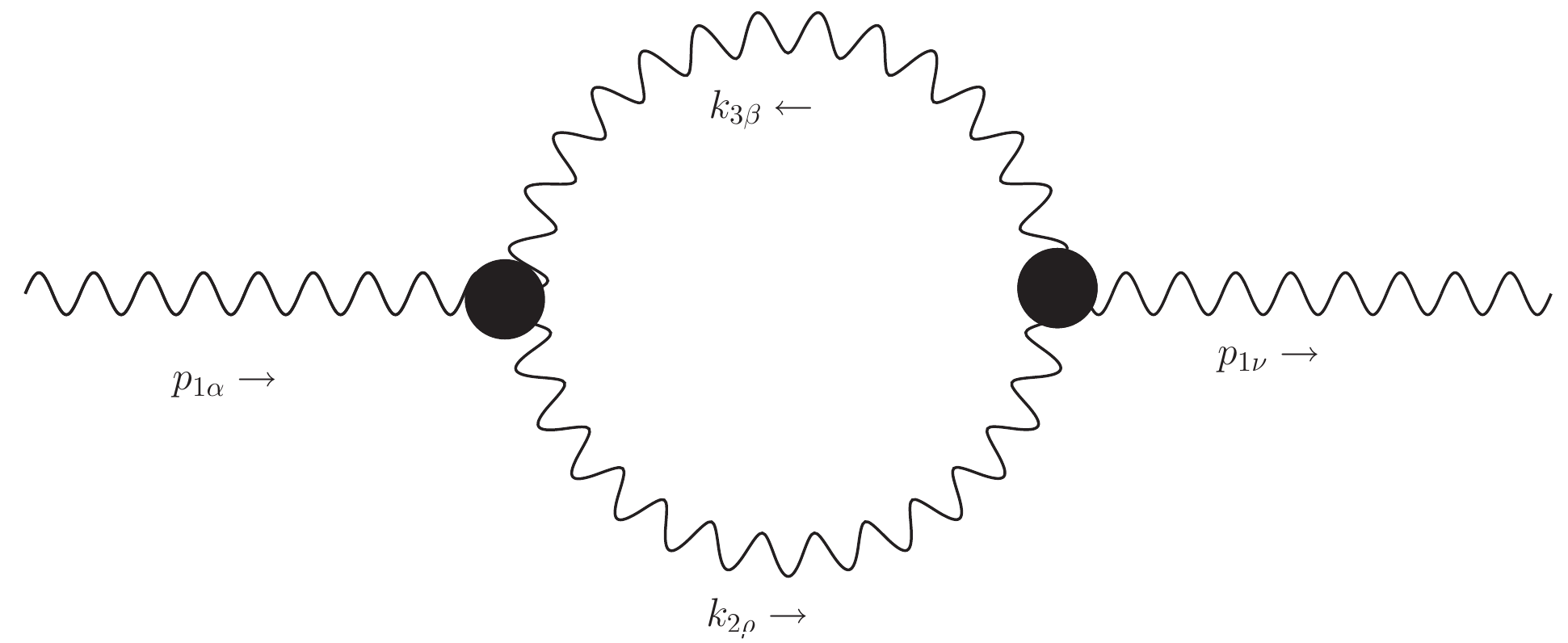}
\caption{\label{propgraph} A one-loop skeleton graph  for   the   proper self-energy.   Blobs represent fully-dressed vertices, and all propagators are fully-dressed.  Other one-loop graphs (ghost, seagull) not shown.   }
\end{center}
\end{figure}
We understand that these graphs are calculated with the gluon propagator of Eq.~(\ref{treeprop}) and a ghost propagator with the same mass, and all the vertices are   tree-level.  There are no explicit $m^2$ terms in any numerator; these appear from seagulls, induced-mass terms, and momentum integrations.

As in the tweaked $\phi^3_6$ model, the divergence in the $1/g_0^2$ Born term will account for the divergence in the proper self-energy.  The SDE we use is:
\begin{equation}
\label{ptself}
  \widehat{\Delta}^{-1}_{\alpha\nu}(p_1)=P_{\alpha\nu}(p_1)\frac{p_1^2}{g_0^2}+\Pi_{\alpha\nu}(p_1)+ \textrm{gauge fixing term}.   
\end{equation}

Begin with the standard PT-RGI expression, with ghosts and gluons having mass $m$:
\begin{eqnarray}
\label{standardprop}
\Pi_{\alpha\nu}(p_2) &  = &  \frac{N}{2(2\pi )^4}\int\!\mathrm{d}^4k_2\{\Gamma_{ \rho\alpha\beta}(-k_2,p_1,k_3)
\Gamma_{\beta\nu\rho}(-k_3,-p_1,k_2)\\ \nonumber
& + &   (k_2+k_3)_{\alpha}(k_2+k_3)_{\nu}\}\frac{1}{(k_2^2+m^2)(k_3^2+m^2)}+\textrm{seagulls}.
\end{eqnarray}
The proper self-energy has been divided by $g_0^2$, and the $\Gamma$s are standard free vertices.

To use the intrinsic pinch, write the vertex numerator as
\begin{equation}
\label{vertexnum}
\Gamma^F\Gamma^F+\Gamma\Gamma^P+\Gamma^P\Gamma-\Gamma^P\Gamma^P.
\end{equation}
 The $\Gamma^F\Gamma^F$ term has no pinches.  We add to the rest the ghost numerator, and
  find
\begin{eqnarray}
\label{pinchterms}
\Gamma\Gamma^P+\Gamma^P\Gamma-\Gamma^P\Gamma^P  & + &  \textrm{ghost numerator}\\ \nonumber
= 4[(p_1^2+m^2)\delta_{\alpha\nu}-p_{1\alpha}p_{1\nu}] & - & \delta_{\alpha\nu}[k_2^2+m^2+k_3^2+m^2]\\ \nonumber
+2(k_2+k_3)_{\alpha}(k_2+k_3)_{\nu} & - & 2m^2\delta_{\alpha\nu}
\end{eqnarray}
where we inserted mass terms.
We have added and subtracted $m^2$ terms to form complete massive (inverse) propagators.  
The first  term in the right-hand side of the equality  is an inverse propagator for the B line of momentum $p_1$ and is dropped (in the PT we can \cite{corn076,cornbinpap} always freely add or drop  any multiple of $p_{1\alpha ,\nu}$).  The second term in the first line gives seagulls, which we do not consider explicitly.  The new term beyond the massless case is the last term.   A similar term arises from integrating the massive propagators with the (rest of the) massless numerator; see Eq.~(\ref{integral2}). 

Doing the algebra, the numerator is that of the massless theory
\begin{equation}
\label{numerator2}
\Gamma^F\Gamma^F=-4(k_2+k_3)_{\alpha}(k_2+k_3)_{\nu}-8(p_1^2\delta_{\alpha\nu}-p_{1\alpha}p_{1\nu})
\end{equation}
plus the {\bf last} line of Eq.~(\ref{pinchterms}).  Some manipulations based on  Sec.~\ref{masspart}, along with seagulls added as needed and the  terms carrying the Nambu-Goldstone-like pole, yield the final PT-RGI proper self-energy:
\begin{equation}
\label{fullprop2}
\Pi_{\alpha\nu}(p_1)  =  \frac{-11NP_{\alpha\nu}(p_1)}{48\pi^4}\int\!\mathrm{d}^4k_2\,\frac{1}{(k_2^2+m^2)(k_3^2+m^2)}[p_1^2-\frac{m^2}{11}]+P_{\mu\nu}(p_1^2)M^2.
\end{equation}
The logarithmic divergence is, as before, cancelled by the bare term in $1/g_0^2$.

\acknowledgments

Many of the results of this paper were first presented at the symposium ``New Physics Within and Beyond the Standard Model", Oberw\"olz, Austria, September 2014.  I thank the organizers for their hospitality at this meeting.

\end{document}